\input{epsf}

\documentstyle[12pt]{article}

\newcommand{\ba}{\begin{array}}
\newcommand{\ea}{\end{array}}
\newcommand{\be}{\begin{equation}}
\newcommand{\ee}{\end{equation}}
\newcommand{\bea}{\begin{eqnarray}}
\newcommand{\eea}{\end{eqnarray}}
\newcommand{\beas}{\begin{eqnarray*}}
\newcommand{\eeas}{\end{eqnarray*}}

\newcommand{\NP}{{\em Nucl.\ Phys.\ }}

\newcommand{\PR}{{\em Phys.\ Rev.\ }}

\newcommand{\PRL}{{\em Phys.\ Rev.\ Lett.\ }}

\font\blackboard=msbm10 at 12pt
\font\blackboards=msbm7
\font\blackboardss=msbm5
\newfam\black
\textfont\black=\blackboard
\scriptfont\black=\blackboards
\scriptscriptfont\black=\blackboardss
\def\bb#1{{\fam\black\relax#1}}

\begin{document}
\pagestyle{plain}
\setcounter{page}{1}

\baselineskip16pt

\begin{titlepage}

\begin{flushright}
PUPT-1803\\
hep-th/9807060
\end{flushright}
\vspace{16 mm}

\begin{center}
{\Large \bf Holomorphic Curves from Matrices}

\vspace{3mm}

\end{center}

\vspace{8 mm}

\begin{center}

Lorenzo Cornalba and Washington Taylor IV\footnote{Address after
August 1998: Center for Theoretical Physics, Massachusetts Institute
of Technology, Cambridge MA 02139; {\tt wati@mit.edu}}

\vspace{3mm}

{\small \sl Department of Physics} \\
{\small \sl Joseph Henry Laboratories} \\
{\small \sl Princeton University} \\
{\small \sl Princeton, New Jersey 08544, U.S.A.} \\
{\small \tt cornalba@princeton.edu, wati@princeton.edu}

\end{center}

\vspace{8mm}

\begin{abstract}

Membranes holomorphically embedded in flat noncompact space are
constructed in terms of the degrees of freedom of an infinite
collection of 0-branes.  To each holomorphic curve we associate
infinite-dimensional matrices which are static solutions to the matrix
theory equations of motion, and which can be interpreted as the matrix
theory representation of the holomorphically embedded membrane.  The
problem of finding such matrix representations can be phrased as a
problem in geometric quantization, where $\epsilon\propto l_P^3/R$
plays the role of the Planck constant and parametrizes families of
solutions.  The concept of Bergman projection is used as a basic tool,
and a local expansion for the action of the projection in inverse
powers of curvature is derived.  This expansion is then used to compute
the required matrices perturbatively in $\epsilon$.  The first two
terms in the expansion correspond to the standard geometric
quantization result and to the result obtained using the metaplectic
correction to geometric quantization.

\end{abstract}

%\vspace{2cm}
\vspace{8mm}
\begin{flushleft}
July 1998
\end{flushleft}
\end{titlepage}
\newpage

%%%%%%%%%%%%%%%%%%%%%%%%%%%%%%%%%%%%%%%%%%%%%%%%%%%%%%%%%%%%%%%%%%%%%%%%%
\section{Introduction}
\label{i}
%%%%%%%%%%%%%%%%%%%%%%%%%%%%%%%%%%%%%%%%%%%%%%%%%%%%%%%%%%%%%%%%%%%%%%%%%

Matrix theory \cite{BFSS} (see
\cite{Bilal-review,Banks-review,Susskind-review,WT-Trieste} for reviews)
has been proposed as a non-perturbative definition of $M$-theory, and
therefore, at low energies, of $11$-dimensional supergravity.  Matrix
theory describes not only fundamental particles like the graviton, but
also extended objects like membranes and $5$-branes.  Matrix theory is
very closely related to the light-front formulation of the
supermembrane, which was originally studied in the thesis of Hoppe
\cite{Goldstone-Hoppe} and in the work of
de Wit, Hoppe and Nicolai \cite{dhn}.  Using this connection one can
describe extended two-dimensional membranes within the context of
matrix theory.  Compact surfaces with spherical or higher-genus
topology were studied in \cite{dhn,Dan-Wati}.  Compact surfaces,
however, are unstable to gravitational collapse, and are expected to
radiate gravitons and to eventually disappear.  On the other hand, if
one focuses attention on static configurations, one is forced to
consider noncompact surfaces which are infinite in spatial extent.
The planar infinite membrane, in particular, has been extensively
discussed in the matrix theory literature \cite{BFSS,bss}.  {}From the
point of view of membrane theory, however, the planar brane is nothing
but a special case of a larger class of static solutions to the
equations of motion, given by holomorphic embeddings of noncompact
Riemann surfaces in space.  These holomorphic membranes are stable,
static configurations corresponding to supergravity solutions which
preserve some supersymmetries and are therefore BPS configurations.
One can ask if there are matrix theory configurations corresponding to
these holomorphic curves in space.  This paper is devoted to analyzing
this question.

We construct in this paper a set of infinite
matrices corresponding to any holomorphic membrane configuration.  To
motivate the general analysis of the paper, we give here an explicit
example of the type of holomorphic matrix membrane in which we are
interested.  Consider a planar membrane embedded in a pair of
holomorphic coordinates $Z = X_1 + i X_2, W  = X_3 + i X_4$ according to
the equation $W = Z^2$.  A static matrix theory configuration
corresponding to this membrane would be a pair of infinite-dimensional
complex matrices ${\cal Z} = {\cal X}_1 + i {\cal X}_2, {\cal W} =
{\cal X}_3 + i {\cal X}_4$ satisfying the
relation ${\cal W} = {\cal Z}^2$ and the equations of motion $[[{\cal
X}_i, {\cal X}_j], {\cal X}_j]= 0$.
Such matrices can be constructed by taking ${\cal W} = {\cal Z}^2$ with
\begin{equation}
{\cal X}_1 =\frac{1}{2}\left(\begin{array}{cccccc}
 0 &  \rho_0 & 0 & 0 & \ddots &\\
\rho_0 & 0 & \rho_1 &0 & 0 & \ddots\\
0 & \rho_1 & 0 &\rho_2 &  0 & \ddots \\
0 & 0 & \rho_2 & 0 & \rho_3 &\ddots\\
\ddots & 0 & 0 & \rho_3 &0 & \ddots\\
& \ddots & \ddots & \ddots &\ddots & \ddots\\
\end{array}\right),
\;\;\;\;\; {\cal X}_2 =\frac{1}{2}\left(\begin{array}{cccccc}
 0 & i \rho_0 & 0 & 0 & \ddots &\\
-i\rho_0 & 0 &i \rho_1 &0 & 0 & \ddots\\
0 &-i \rho_1 & 0 &i\rho_2 &  0 & \ddots \\
0 & 0 &-i \rho_2 & 0 & i\rho_3 &\ddots\\
\ddots & 0 & 0 &-i \rho_3 &0 & \ddots\\
& \ddots & \ddots & \ddots &\ddots & \ddots\\
\end{array}\right)
\label{eq:rhos}
\end{equation}
where the matrix entries $\rho_n$ are given in terms of
$\rho_0 \approx 0.7502, \rho_{-1} = \rho_{-2} = 0$ by the recursive formula
\[
\rho_n = \sqrt{\frac{1 + \rho^2_{n-2}-\rho^2_{n-1} + \rho^2_{n-2}
\rho^2_{n-3}}{ \rho^2_{n-1}} }, \;\;\;\;\; \;\;\; n > 0
\]
We will discuss this example in more detail in section 3.

In this paper we rephrase the problem of finding a matrix
representation of a general holomorphic curve as a problem in
geometric quantization \cite{Woodhouse}.  The constant
$\epsilon\propto l_P^3/R$ (where $l_P$ is the $11$-dimensional Planck
length and $R$ is the light-like compactification radius) plays the
role of the Planck constant $\hbar$, and it parametrizes families of
solutions to the matrix theory equations of motion.  Given a
holomorphic embedding of a Riemann surface $\Sigma$ in space, we wish
to construct matrices which correspond to the embedding coordinates.
The matrices must be infinite-dimensional, reflecting the
noncompactness of the branes, and are therefore operators acting on a
Hilbert space $\cal H$.  After analyzing some examples, we propose
that $\cal H$ be taken to be the space of holomorphic functions on the
Riemann surface $\Sigma$, and that the operators corresponding to
holomorphic functions act via pointwise multiplication.  The problem
of representing holomorphic curves is then reduced to the problem of
finding the correct inner product on $\cal H$.  Different inner
products on $\cal H$ are naturally given by integrations over $\Sigma$
with respect to different volume forms $\Omega$, and therefore the
question about the correct choice of inner product can be rephrased in
terms of the corresponding volume form $\Omega$.  To make this
correspondence we  discuss Bergman projections
\cite{bergman,Tuynman} and kernels on $\Sigma$, and we  derive a
local expansion of the action of the projection.  We then use this
expansion to solve for the volume form $\Omega$, which we do
perturbatively in powers of $\epsilon$.  We then devote the last part
of the paper to describing the connections between our approach and the
theory of geometric quantization.  We find that, to first order in
$\epsilon$, our expression for $\Omega$ reproduces the result expected
from geometric quantization.  The first correction 
corresponds to the metaplectic correction to geometric quantization
\cite{Woodhouse,Tuynman2}.  All higher order terms are needed in order
to satisfy the equations of motion, and cannot be determined from standard
geometric quantization theory.  

The structure of this paper is as follows: in section 2 we review the
light-front description of the classical bosonic membrane and its
connection with matrix theory.  We discuss the static solutions of the
membrane equations of motion corresponding to holomorphically embedded
membranes.  The central problem which we address in this paper, that
of finding matrix representations of holomorphic curves, is defined in
subsection 2.3.  In section 3 we discuss some simple examples of
holomorphic curves corresponding to static membranes; subsection
\ref{sec:example-general} describes an approach to solving the general
problem based on the insights gained from the examples.  Section 4
contains a detailed discussion of Bergman integral kernels.  We review
some basic properties of the projections associated with these
kernels, describe a simple example, and derive a general formula for a
projection operator which agrees with Bergman projection on a very
large class of functions.  In section 5 we use the tools developed in
section 4 to propose a solution to the general problem of constructing
a matrix representation of a holomorphic curve.  Section 6 applies the
general formalism in several examples, including the simple examples
of section 3 as well as more complicated examples which cannot be
solved without the general formalism.  Section 7 contains a discussion
of the connection between the results in this paper and the theory of
geometric quantization.  We conclude in section 8 with suggestions for
future research.

%%%%%%%%%%%%%%%%%%%%%%%%%%%%%%%%%%%%%%%%%%%%%%%%%%%%%%%%%%%%%%%%%%%%%%%%%
\section{Bosonic Membranes and Matrix Theory}

%%%%%%%%%%%%%%%%%%%%%%%%%%%%%%%%%%%%%%%%%%%%%%%%%%%%%%%%%%%%%%%%%%%%%%%%%

\subsection{The light-front bosonic membrane and holomorphic curves}
\label{m1}

In this subsection we briefly review the theory of a classical bosonic
membrane moving in 11 dimensions in light-front coordinates.  The
derivation of the light-front formalism starting from the Nambu-Goto
action has been extensively discussed in the literature, and we refer
the reader to the articles \cite{Goldstone-Hoppe,dhn} for a more
detailed explanation.  At the end of this subsection we discuss the
particular class of static solutions of the equations of motion for
the brane that is the central focus of the rest of the paper.

We take the world-volume of the membrane to be the product space ${\bb
R}\times \Sigma$, where $\Sigma$ is a two dimensional surface (not
necessarily compact) with the topology of the brane.  Coordinates on
the world-volume are $\tau$ and $\sigma^a$, with $a=1,2$.  The brane
propagates in $11$-dimensional Minkowski space-time with
coordinates $X^\mu$, $\mu=0 \dots 10$, and the motion of the surface
is described by coordinate functions $X^\mu : {\bb R}\times \Sigma
\rightarrow {\bb R}$ on the world-volume.  We  use light-front
coordinates $X^{\pm}$ given by
\bea
X^{\pm}&=&{1\over\sqrt{2}} (X^0 \pm X^{10})\nonumber\\
X_{\pm}&=& -X^{\mp}={1\over\sqrt{2}} (X_0 \pm X^{10}).\nonumber
\eea
The light-front formalism is based on a simple observation.  The
membrane, as it moves in space-time, carries a conserved momentum
$p^\mu$, which can be written, as always in field theory, as an
integral over $\Sigma$ of densities $P^\mu$
$$
p^\mu = \int_\Sigma P^\mu d^2\sigma.
$$
In general the densities $P^\mu$ will depend on $\tau$.  On the other
hand, if one imposes the light-front gauge constraints (to be
described shortly), one can show that $P^+$ is independent of $\tau$.
Therefore $P^+$ singles out, up to a multiplicative constant, a fixed
volume form on $\Sigma$ which we call
$$
\mu\ d^2\sigma.
$$
We may then write the momentum densities $P^\mu$ as
\bea
P^+ &=& \Pi^+ \mu \nonumber\\
P^- &=& \Pi^- \mu \nonumber\\
P^i &=& \Pi^i \mu, \hskip1cm(i=1\dots 9)\nonumber
\eea
where $\Pi^+$ is a fixed constant and $\Pi^-,\Pi^i$ are scalar
functions on $\Sigma$.  The hamiltonian light-front formalism starts
then with the choice of a fixed volume form $\mu$ and takes as
independent variables the transverse positions $X^i$ together with the
corresponding momenta $\Pi^i$, subject to one constraint.  The
longitudinal coordinates $X^0$ and $X^{10}$ are then derived by
solving the gauge-fixing equations.

Let us first describe the hamiltonian which governs the evolution of the 
independent canonical variables, together with the constraint satisfied by them.  To this
end we view $\mu$ as a symplectic form and we consider the corresponding Poisson bracket.
Specifically, if $A$ and $B$ are functions on $\Sigma$, we define the bracket $\{A,B\}$
by
$$
\{A,B\} = {1\over\mu} \epsilon^{ab} \partial_a A\partial_b B,
$$
where $\epsilon^{12}=1$.  With this notation we can then write the hamiltonian
describing the membrane (T is the brane tension)

\bea\label{m1.15}
H &=& \int_\Sigma \mu\ d^2\sigma\ \cal H \\
\cal H &=& {1\over 2\Pi^+} \Pi^i\Pi^i + {T^2\over 4\Pi^+} \{X^i,X^j\}^2\nonumber
\eea
together with the constraint satisfied by the canonical variables

\be\label{m1.1}
\{X^i,\Pi^i\}=0.
\ee
The second term in the hamiltonian can also be rewritten in terms of the 
induced metric $$h_{ab}=\partial_a X^i \partial_b X^i$$ on $\Sigma$
by noticing that $$h=\det\ h_{ab}= {1\over 2} \mu^2  \{X^i,X^j\}^2.$$
It is easy to show, using the canonical commutation relation
$$
[X^i(\sigma), \Pi^i(\sigma ')]_{P.B.} = {1\over\mu(\sigma)} \delta^2(\sigma-\sigma '),
$$
that the equations of motion are given by

\bea\label{m1.2}
\Pi^i &=& \Pi^+ \dot X^i\nonumber\\
\ddot X^i &=& {1\over \Pi^+} \dot{\Pi}^i = {T^2\over {\Pi^+}^2} \{\{X^i,X^j\},X^j\}.
\eea
A simple application of the Jacobi identity then shows that the constraint
(\ref{m1.1}) is preserved by the hamiltonian evolution (\ref{m1.2}) and that
the hamiltonian system at hand is consistent.

We now turn our attention to the gauge  fixing equations that determine the
constraint coordinates $X^+$ and $X^-$.  They read
\bea\label{m1.3}
X^+ &=& \tau\nonumber\\
\partial_a X^- &=& \dot X^i \partial_a X^i.
\eea
The first equation simply says that the hamiltonian time $\tau$ measures the light-front
coordinate $X^+$.  Since the momenta conjugate to $\tau$ and $X^+$ are respectively $H$ and
$p^-$, one has that $${\cal H} = \Pi^-.$$ The second equation can be solved for $X^-$, at least
locally, if the right hand side of equation (\ref{m1.3}) is closed.  But this is the case since
$
d(\dot X^i dX^i) \propto d(\Pi^i dX^i) = d\Pi^i\wedge dX^i =
 \epsilon^{ab}\partial_a \Pi^i\partial_b X^i\ d^2\sigma= \{\Pi^i,X^i\} \mu d^2\sigma = 0
$.
Global problems of existence of $X^-$ will not be an issue in this paper and we will not 
address them.  

To end our discussion of the light-front formalism, we derive an equation for the density $\mu$
in terms of the coordinate functions $X^\mu$.  We start by taking the time derivative of the gauge
constraint (\ref{m1.3}).  This can be rewritten as $\partial_a \Delta = 2\ddot X^i\partial_a X^i=
2(T^2/{\Pi^+}^2) \{\{X^i,X^j\},X^j\}\partial_a X^i$, where
$$
\Delta = 2\dot X^- - \dot X^i \dot X^i.
$$
A mechanical computation shows that $4\{\{X^i,X^j\},X^j\}\partial_a X^i = \partial_a \{X^i,X^j\}^2 =
2\partial_a (h/\mu^2)$.  Putting everything together we deduce that 
$$
\mu= {T\over\Pi^+}\sqrt{h\over\Delta}.
$$

We conclude this section by focusing our attention on a specific class
of static solutions of the equations of motion.  We first of all fix,
on the space part of Minkowski space-time, a complex structure
compatible with the metric.  All the possible choices differ only by
an $SO(10)$ rotation; we choose the analytic
coordinates
\bea
Z_1 &=& X^1+iX^2\nonumber\\
&&\dots\nonumber\\
Z_5 &=& X^9+iX^{10}.\nonumber
\eea
We then choose a complex structure on the manifold $\Sigma$, denoting
the analytic coordinate  by $$z=\sigma^1 + i\sigma^2.$$ 
A class of static
solutions of the equations of motion are then given by holomorphic
embeddings of $\Sigma$ in ${\bb R}^{10} = {\bb C}^5$.  More
specifically we shall take
\bea\label{m1.4}
Z_A &\in &\{ {\rm analytic\ functions\ on}\; \Sigma\ (\tau\
{\rm independent}) \}\nonumber\\ 
Z_5 &= &0\\ X^+ &=&X^-= \tau.\nonumber
\eea
The surface $\Sigma$ cannot be compact.  If it were, then the
embedding would be trivial since holomorphic functions on a compact
Riemann surface are constant, so the brane would degenerate to a
point.  To prove that the equations of motions are satisfied for a
holomorphically embedded membrane, we first
note that the gauge fixing equations hold.  We then compute the
induced metric $h_{ab}$, which is given by
\bea
h_{zz} &=& h_{\bar z\bar z}=0\nonumber\\
h_{z\bar z} &=& {1\over 2} \partial Z_A\bar\partial \bar Z_A\nonumber
\eea 
(sum over $A$ will always be implied).  Using the fact that 
\bea
\Delta &=& 2\nonumber\\
\sqrt h &=& 2 h_{z\bar z} \nonumber
\eea
we can then compute the density $\mu$ given by
$$
\mu = \left( {T\over \sqrt 2\Pi^+} \right) \partial Z_A\bar\partial \bar Z_A = {1\over \pi\epsilon} \partial Z_A \bar
\partial \bar Z_A,
$$
where
$$
\epsilon = 2\sqrt{2} {\Pi^+\over 2\pi T}.
$$
The symplectic form $\mu\ d\sigma^1\wedge d\sigma^2$ can be rewritten
as ${i\over 2} \mu\ dz\wedge d\bar z$ and correspondingly the bracket
$\{ , \}$ can be expressed in terms of holomorphic and antiholomorphic
derivatives as $$\{ A,B \} = {-2i\over\mu} (\partial A\bar\partial
B-\bar\partial A \partial B).$$ It is then easy to show that
$$ \{ Z_A, Z_B \} = \{ \bar Z_A, \bar Z_B \} = 0$$
and also that
\be\label{m1.5}
\{ Z_A, \bar Z_A \} = {-2i\over \mu} \partial Z_A \bar\partial Z_A = -2\pi i\epsilon.
\ee
We will see later the significance of the constant $\epsilon$.  For now we can use the above results together with
the Jacobi identity to show that 
\bea\label{m1.14}
\ddot Z_A & \propto & \{ \{ Z_A,Z_B \} , \bar Z_B \} +  \{ \{ Z_A,\bar Z_B \} , Z_B \} 
=  \{ \{ Z_A, \bar Z_B \} , Z_B \} = \nonumber\\
&=&  \{ \{ Z_B,\bar Z_B \} , Z_A \} +  \{ \{ Z_A,Z_B \} , \bar Z_B \}= \{ -2\pi i\epsilon,Z_A \} = 0.
\eea
The equations of motion are therefore satisfied and we indeed have a
static solution of the hamiltonian equations (\ref{m1.2}).  The
solution (\ref{m1.4}) can also be boosted in the $10$-th direction.
If $\omega$ is a boost parameter, then
\bea
X^+ \rightarrow \omega X^+ &=& \omega \tau\nonumber\\
X^- \rightarrow {1\over\omega} X^- &=& {1\over\omega} \tau.\nonumber
\eea
The above transformation does not preserve the gauge condition, and we have to rescale $\tau\rightarrow{1\over\omega}\tau$.
We than have that
$$
X^+=\tau\hskip1cm X^- = {1\over\omega^2}\tau.
$$
The only change in the above discussion is that $\Delta\rightarrow {2\over\omega^2}$ and therefore $\mu\rightarrow\omega\mu$.
This is then reflected in a change
$$
\epsilon\rightarrow{1\over\omega}\epsilon
$$
in equation (\ref{m1.5}).

%%%%%%%%%%%%%%%%%%%%%%%%%%%%%%%%%%%%%%%%%%%%%%%%%%%%%%%%%%%%%%%%%%%%%%%%%
\subsection{Matrix-membrane correspondence}
\label{m2}
%%%%%%%%%%%%%%%%%%%%%%%%%%%%%%%%%%%%%%%%%%%%%%%%%%%%%%%%%%%%%%%%%%%%%%%%%

We have briefly reviewed the theory of classical membranes, with
attention to a particular class of static solutions.  In this
subsection we discuss the relation between light-front membrane theory
and matrix theory.

It was pointed out in \cite{Goldstone-Hoppe,dhn} that the light-front
membrane theory discussed in the previous section can be related to a
theory of matrices by truncating the space of functions on the brane
to a finite number of degrees of freedom.  This gives a discrete
regularization of the membrane theory which preserves much of the
structure of the continuous theory.  The matrix theory conjecture of
Banks, Fischler, Shenker and Susskind, first proposed in \cite{BFSS}
and then further developed in
\cite{Susskind-DLCQ,Sen-DLCQ,Seiberg-DLCQ} (for reviews see
\cite{Bilal-review,Banks-review,Susskind-review,WT-Trieste}),
asserts that the supersymmetric version of this matrix quantum
mechanics theory contains all the physics of light-front M-theory.
More precisely, the DLCQ version of the conjecture
asserts that M-theory compactified on a light-like circle
$X^- \sim X^- + 2\pi R$ is described, within the sector with
light-front momentum $p^+=N/R$, by $10$-dimensional $U(N)$ super
Yang-Mills theory, dimensionally reduced to $0+1$ dimensions.  
The supermembrane of M-theory is described in matrix theory using
precisely the matrix-membrane correspondence worked out by de Wit,
Hoppe and Nicolai in \cite{dhn}.

Before discussing the details of the matrix-membrane correspondence,
let us fix some conventions.  The $11$-dimensional Planck length is
denoted by $l_P$ and is related to the gravitational constant by
$2\kappa^2=(2\pi)^8 l_P^9$.  The membrane tension $T$ is 
$$T={1\over (2\pi)^2 l_P^3}.$$ Finally the string scale is given by
$\alpha^\prime = l_P^3/R$.

We now move to an overview of matrix theory.  The independent
variables are given by the transverse coordinates $X^i$ together with
the corresponding canonical momenta $\Pi^i$ ($i=1\dots 9$), where now
both $X^i$ and $\Pi^i$ are $N\times N$ hermitian matrices.  The
canonical variables are not completely independent but satisfy a
constraint equation given by
\be\label{m2.1}
[X^i,\Pi^i]=0.
\ee
Time evolution is governed by the Hamiltonian
\be\label{m2.5}
H={R\over 2} {\rm Tr}\;(\Pi^i \Pi^i) - (2\pi T)^2 {R\over 4} {\rm Tr}\;([X^i,X^j]^2)
\ee
from which the equations of motion
\bea\label{m2.2}
\Pi^i &=& {1\over R}\dot X^i\\
\ddot X^i &=& R \dot\Pi^i = -(2\pi)^2 T^2 R^2 [[X^i,X^j],X^j]\nonumber
\eea
can be derived.  The constraint (\ref{m2.1}) is preserved by
(\ref{m2.2}) and therefore the theory is consistent.  In the
$11$-dimensional interpretation of matrix theory the conserved
momentum $p^\mu$ is given by
\bea
p^+ &=& {N\over R}\nonumber\\
p^- &=& H \nonumber\\
p^i &=& {\rm Tr}\;(\Pi^i).\nonumber
\eea
There is an obvious formal similarity between this matrix quantum
mechanics theory and the membrane theory reviewed in the previous
section.  This connection was made precise in
\cite{Goldstone-Hoppe,dhn}.  A configuration of a
membrane $\Sigma$ can be associated
with a set of $N \times N$ matrices by mapping functions on the
membrane, like coordinates and momenta, into $N \times N$ matrices in
matrix theory.  Through this correspondence, Poisson brackets $\{ ,
\}$ on $\Sigma$ become matrix commutators $[,]$ and integrations $\int_\Sigma
d^2\sigma\,\mu$ are replaced by traces ${\rm Tr}$ of matrices.  The
map from functions on $\Sigma$ to matrices was described in detail in
\cite{dhn} in the cases where $\Sigma$ is a Riemann surface of
spherical or toroidal topology.  We wish to discuss this
correspondence in the more general case where the membrane $\Sigma$ is
a noncompact Riemann surface.  We conclude this subsection by
reviewing the situation when $\Sigma$ is compact.

To make the matrix-membrane correspondence more
precise, let $\cal A$ be the space of scalar functions on $\Sigma$ and
$\cal B$ be the space of $N\times N$ matrices.  Both spaces $\cal A$
and $\cal B$ carry a similar algebraic structure and we should
therefore look for a correspondence ${\cal Q}:{\cal A}
\rightarrow {\cal B}$ which preserves this structure as much as
possible.  Obviously $\cal Q$ should be a linear map and should map
complex conjugate functions to hermitian conjugate matrices $$\bar X
\rightarrow X^\dagger.$$  For some functions on $\Sigma$ we wish the
product of functions to correspond to a matrix product in ${\cal B}$.
This is not possible in general, since the matrix product is not
commutative.  On the other hand, one can at least require that the
unit element in $\cal A$ be mapped to the unit element in $\cal B$
\be\label{m2.3}
1\rightarrow {\bf 1}_{N\times N}.
\ee
We recall from the last section that the measure $\mu$ is defined only up to a multiplicative factor
(the product $\Pi^+ \mu = P^+$ is invariant and we can rescale $\mu$ as long as we rescale 
$\Pi^+$ accordingly).  We use this freedom to fix the correspondence
\be\label{m2.4}
\int_\Sigma d^2\sigma\ \mu \rightarrow {\rm Tr}\;(\ ).
\ee
Combining (\ref{m2.3}) and (\ref{m2.4}) we see that the normalization
of $\mu$ has been chosen so that
$$\int_\Sigma d^2\sigma\ \mu = N.$$ 
In the language of matrix theory, the measure $\mu$ corresponds to
the local density of 0-branes on the membrane, and $N$ is the total
number of 0-branes.
If we recall that $p^+ = N/R =
\int_\Sigma d^2\sigma\ \mu\Pi^+$, we also conclude that $$\Pi^+ =
{1\over R}.$$ The last requirement on $\cal Q$ comes from the fact
that both $\cal A$ and $\cal B$ are Lie algebras, with bracket $\{ ,
\}$ and $[,]$ respectively.  In order to match normalizations in the
hamiltonians (\ref{m1.15}) and (\ref{m2.5}), we require that, under
$\cal Q$,
\be\label{m2.10}
\{\, ,\, \} \rightarrow 2\pi i\ [\, ,\, ].
\ee 
This is clearly an example of the classical problem of geometric
quantization, if one views $\Sigma$ as a symplectic manifold with
symplectic form $\mu\,d^2\sigma$.  We shall see later that we will not
be able to satisfy (\ref{m2.10}) for all elements of $\cal A$ (this is
reminiscent of similar problems in elementary quantum mechanics).  On
the other hand we will see that, in the cases that we shall study, it
is possible and natural to impose (\ref{m2.10}) on the coordinate
functions describing the position of the brane.

This concludes our general discussion of matrix theory and of its
correspondence with membrane theory described in the last section.  We
are now in a position to present clearly the problem that will be
analyzed in this paper.  

\subsection{Holomorphic curves in matrix theory}

At the end of section \ref{m1} we
described a family of static solutions of the membrane equations of
motion, given by holomorphic embeddings of a Riemann surface $\Sigma$
in ${\bb C}^4$ (we had chosen $Z_5=0$).  As we noted already, $\Sigma$
cannot be compact and we therefore choose $\Sigma$ to be a Riemann
surface of genus $g$ with $n$ points deleted.  Moreover we choose
the coordinate functions $Z_A$ ($A=1\dots 4$) to be holomorphic
functions on $\Sigma$, meromorphic at the punctures.  The measure
$\mu$ is given by $\mu=(1/\pi\epsilon)\partial Z_A \bar\partial \bar
Z_A$.  
Since $N=\int_\Sigma d^2\sigma\ \mu=\infty$, we have to change
our point of view slightly, and let ${\cal B}=End({\cal H})$ be the
space of operators acting on an infinite-dimensional Hilbert space
$\cal H$.  Our problem will therefore be, given $\Sigma$ and
holomorphic functions
$Z_A$, to
find a Hilbert space $\cal H$ and a map ${\cal Q}:{\cal A}
\rightarrow End({\cal H})$ satisfying the requirements described
above.   Note that the space ${\cal H}$ and the map ${\cal
Q}$ depend on 
the choice of surface $\Sigma$ and the embedding functions $Z_A$.

We now state in detail the properties which must be satisfied by the
map ${\cal Q}$ for a
quantization of the holomorphic membrane embedding given by a set of
functions $Z_A$.  If
$${\cal Z}_A = {\cal Q}(Z_A)$$ we will require that 
\be\label{m2.6}
[{\cal Z}_A, {\cal Z}_B]= [{\cal Z}_A^\dagger, {\cal Z}_B^\dagger] = 0
\ee
and that
\be\label{m2.8}
[{\cal Z}_A, {\cal Z}_A^\dagger]=-\epsilon \;{\rm Id},
\ee
where
$$
\epsilon = {2\sqrt{2}\Pi^+\over 2\pi T}= 2\sqrt{2} {1\over R} (2\pi l_P^3)=2\sqrt 2 (2\pi\alpha^\prime).
$$
It is clear that the constant $\epsilon$ plays the role of the Planck
constant in the usual classical/quantum correspondence, and that it
will parametrize a family of solutions to the quantization problem.
For this reason in the rest of the paper we will loosely refer to
$\epsilon$ as the Planck constant, to the limit $\epsilon\rightarrow
0$ as the classical limit and to the process of going from functions
to matrices through $\cal Q$ as quantization.  Let us note that one
can use equations (\ref{m2.6}) and (\ref{m2.8}), together with a
manipulation identical to equation (\ref{m1.14}) to show, starting
from the equations of motion, that
$$\ddot{\cal Z}_A = 0.$$ Therefore the matrices $\cal Z_A$ represent a
static solution of the matrix theory equations of motion.

Equation (\ref{m2.6}) allows us to impose one final and crucial
constraint on $\cal Q$.  Suppose that the embedded surface $\Sigma
\hookrightarrow {\bb C}^4$ can be described as the locus of points in
${\bb C}^4$ satisfying the equations
\be\label{m2.7}
F_i(Z_A) = 0 \hskip1cm (i=1\dots 3)
\ee
for some holomorphic functions $F_i$ (the functions $F_i$ could be polynomial functions of the $Z_A$,
but this is not necessary).  Since the operators ${\cal Z}_A$ commute, it makes sense to replace 
$Z_A$ with ${\cal Z}_A$ in equation (\ref{m2.7}) and to require that 
\be\label{m2.9}
F_i({\cal Z}_A) = 0.  \hskip1cm (i=1\dots 3)
\ee

Solutions to equations (\ref{m2.6}), (\ref{m2.8}) and (\ref{m2.9})
will not be unique, since the underlying theory is invariant under
$U(N)$ gauge transformations.  In particular we are free to transform
the operators ${\cal Z}_A$ to $U {\cal Z}_A U^\dagger$, where $U$ is a
unitary matrix.

As a final remark let us discuss boosted solutions.  At the end of the
last section we noted that a boost in the $10$-th direction with
parameter $\omega$ is only reflected in the change
$$\epsilon\rightarrow {1\over\omega}\epsilon = {2\sqrt 2\over \omega R} (2\pi l_p^3).$$
This means that a static solution of equations (\ref{m2.6}), (\ref{m2.8}) and (\ref{m2.9})
with light-like radius $R$ is equivalent to a boosted solution with light-like radius $R/\omega$.  This
is consistent since, under a boost, $X^-\rightarrow X^-/\omega$ and therefore $R\omega$ is invariant.

%%%%%%%%%%%%%%%%%%%%%%%%%%%%%%%%%%%%%%%%%%%%%%%%%%%%%%%%%%%%%%%%%%%%%%%%%
\section{Simple Examples of Holomorphic Membranes}
\label{m3}
%%%%%%%%%%%%%%%%%%%%%%%%%%%%%%%%%%%%%%%%%%%%%%%%%%%%%%%%%%%%%%%%%%%%%%%%%

In subsection 2.3 we described the general problem of constructing a
Hilbert space ${\cal H}$ and a map ${\cal Q}$ from functions on a
membrane $\Sigma$ to matrices which would give a general matrix
representation of holomorphic curves.  In this section we analyze a few
special examples of holomorphic curves where there is a simple and
natural choice of $\cal H$ and $\cal Q$.  The purpose in discussing
these examples is twofold.  First of all, these examples can be solved
without resorting to the general machinery developed in the rest of
the paper, and are therefore interesting in their own right.
Moreover, the analysis of these examples will suggest a solution to
the general problem, which is discussed in subsection
\ref{sec:example-general} and is the subject of the rest of the paper.

All of the examples that we discuss in this section have a common
underlying structure.  We take $\Sigma$ to be the full complex plane
$\bb C$, with analytic coordinate $z$, and we look for coordinates $x$
and $y$ on $\Sigma$ such that the symplectic form $\mu\,d^2\sigma$
takes the simple form $$\mu\,d^2\sigma = {1\over\pi\epsilon} dx\wedge
dy.$$ We may then consider $x$ and $y$ to be standard canonical
coordinates in a $2$-dimensional phase space and we can perform
quantization similarly to elementary quantum mechanics.  It is 
convenient to define the complex coordinate $$s=x+iy$$ even though, as
will be clear from the examples, $s$ is not necessarily an analytic
coordinate on $\Sigma$ -- i.e.   is not necessarily an analytic function of
$z$.  In terms of $s$ the symplectic form is given by
\be\label{m3.1}
\mu\,d^2\sigma = -{1\over 2\pi i} {1\over \epsilon} ds\wedge d\bar s
\ee
so that $$\{ s,\bar s\} = - 2\pi i\epsilon.$$ Since the symplectic
form is canonical in terms of the coordinate $s$, it is easier to
define the quantization map ${\cal Q}$ on $s$ than directly on the
functions $z$ or $Z_A$.  If we define $$a = {1\over\sqrt\epsilon}
{\cal Q}(\bar s)$$ we conclude that
$$[a,a^\dagger] = 1.$$ The above is a canonical creation-annihilation
pair and the quantization is standard.  The Hilbert space $\cal H$ on
which operators act is spanned by the simple harmonic oscillator
states $|n\rangle$ ($n=0,1,\dots$).  Moreover functions of $s$ and
$\bar s$ are quantized using the correspondence
\bea
s &\rightarrow & \sqrt\epsilon a^\dagger\nonumber\\
\bar s &\rightarrow & \sqrt\epsilon a.\nonumber
\eea 
This map is not uniquely defined for a general function of $s$ and
$\bar{s}$ due to operator ordering ambiguities.  On the other hand
these problems can be resolved by imposing equations (\ref{m2.6}),
(\ref{m2.8}) and (\ref{m2.9}) as conditions on the operators ${\cal
Z}_A ={\cal Q} (Z_A)$.

\subsection{Example: flat membrane}

We now move on to the explicit examples.  The first one is very simple
and well-known.  We wish to describe a flat membrane stretched in the
$1-2$ plane, defined by
\bea
Z_1 &=& z\nonumber\\
Z_A &=& 0.\hskip1cm (A=2,3,4)\nonumber
\eea
Since $$\mu\,d^2\sigma = -{1\over 2\pi i} {1\over \epsilon} dz\wedge d\bar z,$$ we may choose
$s=z$.  Quantization is therefore trivial and is given by 
\bea
{\cal Z}_1 &=& \sqrt\epsilon a^\dagger\nonumber\\
{\cal Z}_A &=& 0.\hskip1cm (A=2,3,4)\nonumber
\eea
The operator $a^{\dagger}$ can be written in terms of hermitian
operators $q, p$ satisfying $[q, p] = i$ through $a^{\dagger} =
(q-ip)/\sqrt{2}$.  This corresponds to a description of the flat
membrane in terms of the hermitian matrices
\begin{eqnarray*}
{\cal X}_1 & = &  \frac{\sqrt{\epsilon}}{\sqrt{2}} q\\
{\cal X}_2 & = & - \frac{\sqrt{\epsilon}}{\sqrt{2}} p
\end{eqnarray*}
exactly as discussed in \cite{BFSS}.  Note that the space of
flat membrane solutions is parameterized by $\epsilon$, which is the
inverse of the 0-brane density $\mu$ on the brane.

\subsection{Example: parabolic membrane}

Let us move to a more complicated example, namely that considered in
the introduction.  In this case we wish to
describe a parabolic surface defined by
\bea
Z_1^2 &=& \beta Z_2\nonumber\\
Z_3 &=& Z_4=0.\nonumber
\eea
where $\beta$ is a numerical coefficient which we take to be 1 for
simplicity.  If we parametrize the surface through
\bea
Z_1 &=& z\nonumber\\
Z_2 &=& z^2,\nonumber
\eea
we can then easily compute
$$ \mu\,d^2\sigma = - {1\over 2\pi i\epsilon} (1+4z\bar z)dz\wedge d\bar z.$$
We now look for a change of coordinates $z\rightarrow s$ so that equation (\ref{m3.1})
is satisfied.  We therefore require that 
\be\label{m3.2}
(1+4z\bar z)dz\wedge d\bar z = ds\wedge d\bar s.
\ee
Both sides of the above equation define rotationally invariant
measures on the plane (under rotations $z\rightarrow z\,e^{i\theta}$
and $s\rightarrow s\,e^{i\theta}$ respectively).  We may then assume
that $z$ is just a radius-dependent rescaling of $s$, or more
precisely that
\be\label{m3.3}
z=sf(s\bar s),
\ee
where $f$ is a positive real function.  In fact one may use
(\ref{m3.3}) to show that equation (\ref{m3.2}) is satisfied provided
that ${dg\over dx} = 1/(1+4g)$, where $g(x)=xf^2(x)$.  On the other
hand we will not need the precise form of $f$ in the sequel, as will
be clear shortly.

Quantization of equation (\ref{m3.3}) by realizing ${\cal Z} ={\cal Q}
(z)$ as an operator on ${\cal H}$ is ambiguous due to
operator-ordering problems.  On the other hand, regardless of the
ordering prescription chosen, the operator $$\rho={\cal Q}(z)$$ must
contain one more creation operator then it contains annihilation ones.
More formally, if $N=a^\dagger a$ is the number operator, then
\be\label{m3.7}
[N,\rho]= \rho.
\ee
Therefore
\bea\label{m3.6}
\rho |n\rangle &=& \rho_n |n+1\rangle \\
\rho^\dagger |n\rangle &=& \bar\rho_{n-1} |n-1\rangle\nonumber
\eea
for some complex coefficients $\rho_n$ $(n\ge 0)$.  If we choose
\bea
{\cal Z}_1 &=& \rho\nonumber\\
{\cal Z}_2 &=& \rho^2,\nonumber
\eea
then equations (\ref{m2.6}) and (\ref{m2.9}) are automatically satisfied.  We
will just have to impose equation (\ref{m2.8}), which reads in this case
\be\label{m3.4}
[\rho,\rho^\dagger]+[\rho^2,{\rho^\dagger}^2]=-\epsilon.
\ee
The above equation will determine the coefficients $\rho_n$ up to a
phase factor.  This is expected since phases can be changed with a
$U(N)$ gauge transformation $\rho\rightarrow U\rho U^\dagger$.  Acting
with equation (\ref{m3.4}) on the state $|n\rangle$ and defining the
positive coefficients $$\alpha_n= | \rho_n |^2,$$ we arrive at the
recursion relation
\be\label{m3.5}
\alpha_n - \alpha_{n-1}+\alpha_{n+1}\alpha_n - \alpha_{n-1}\alpha_{n-2} = \epsilon
\ee
where we define $\alpha_n=0$ for $n<0$.  Equation (\ref{m3.5})
determines $\alpha_{n+1}$ as a rational function of
$\alpha_{n-2}$,$\alpha_{n-1}$,$\alpha_n$.  The only undetermined
coefficient is $\alpha_0$ and if we set $$\alpha_0=\xi$$ then all of
the coefficients $\alpha_n=\alpha_n(\xi)$ are rational functions of
$\xi$ alone.  It seems as though we have thus constructed, for a fixed
$\epsilon$, a family of solutions parametrized by $\xi$.  On the other
hand we note that the $\alpha_n$ are positive coefficients and, for a
generic choice of $\xi$, some of the $\alpha_n(\xi)$ generated by the
recursion relation (\ref{m3.5}) are negative.  Let us denote by
$\Gamma_N\subset {\bb R}$ the subset of the possible values of $\xi$
for which $\alpha_n\ge 0$ for $n\le N$.  Clearly $\Gamma_N\subset
\Gamma_M$ for $N\ge M$.  Moreover one can show numerically that the
$\Gamma_N$'s are nested rectangles which converge rapidly to a unique
value for $\xi$, which can be computed to arbitrary accuracy by taking
$N$ to be large enough.  As a function of $\epsilon$, Figure $1$ (at
the end of section \ref{n}) shows the value of $\alpha_0=\xi$
determined using this procedure (the curve labeled ``exact result'' in the
figure).  The example discussed in the introduction corresponds to
taking $\epsilon = 1$.
Although it is clearly indicated by the numerics, we do not have a
proof that $\alpha_0$ is uniquely determined by this procedure; we
will be content here with the numerical result, however, because later
we give a more general algebraic procedure which will include
this special example.

We have thus found a matrix representation of the holomorphic curve
$Z_2= Z_1^2$ for any value of the parameter $\epsilon$.  As in the
example of the flat membrane in the previous section, $\epsilon$ is
related to the inverse of the 0-brane density on the membrane.  A
physical picture of how the 0-branes are distributed on the surface of
the parabolic membrane can be obtained by considering the matrix
\[
{\cal X}_1^2 +{\cal X}_2^2 =\frac{1}{2} (\rho \rho^{\dagger} +
\rho^{\dagger} \rho) = {\rm Diag} (\frac{\alpha_0}{2},
\frac{\alpha_0 + \alpha_1}{2},
\frac{\alpha_1 + \alpha_2}{2},
\frac{\alpha_2 + \alpha_3}{2},
\ldots )
\]
Because this matrix is diagonal, we can think of the individual
0-branes as having well-defined values of $r^2 = x_1^2 + x_2^2$.
Thus, the individual 0-branes are in a sense localized on circular
orbits of radii 
\begin{equation}
r_n = \sqrt{(\alpha_{ n-1} + \alpha_n)/2}.
\label{eq:radii}
\end{equation}  
We can
compare this picture to the expectation that the 0-branes are
uniformly distributed on the membrane surface.  For large $N$, $r_N
\sim \rho_N$.
From (\ref{eq:radii})
we see that as $N$ becomes large, we expect to have $N$ 0-branes
distributed over the portion of the membrane with $r^2 < \alpha_N$.
The area of this portion of the membrane is given by
\[
A = {1\over \epsilon} \int_0^r (1 + 4 t^2) 2 t d t = {1\over \epsilon} (2 r^4 + r^2).
\]
Setting $A = N$ gives
\[
r^2 = \frac{\sqrt{8 \epsilon N + 1}-1}{4}  \sim
\frac{ \sqrt{\epsilon N}}{ \sqrt{2}}  -\frac{1}{4}  +{\cal O} (N^{-1/2}).
\]
On the other hand,
from the recursion relation (\ref{m3.5}) we can derive the relation
\[
\alpha_{N-1} (\alpha_N + \alpha_{N-2}+ 1) = \epsilon N
\]
from which we can determine the asymptotic form of $\alpha_N$
\[
\alpha_N \sim \frac{ \sqrt{\epsilon N}}{ \sqrt{2}}  -\frac{1}{4}  +{\cal O} (N^{-1/2}).
\]
This shows that the simple physical picture of the 0-branes being
localized on circles of radii $\rho_N$ is quite accurate as $N
\rightarrow \infty$.

\subsection{Example: general rotationally invariant curve}

We can generalize the previous example with very little effort as
follows (the steps are identical and we will be brief).  We consider a
surface defined by $$Z_A = c_A\cdot z^{p_A},$$ where the $p_A$
positive integers with no common divisor and the $c_A$ are complex
coefficients.  The volume form on $\Sigma$ is then $$\mu\,d^2\sigma =
-{1\over 2\pi i
\epsilon}\left(\sum_A |c_A|^2 p_A^2 (z\bar z)^{p_A-1}\right) dz\wedge
d\bar z$$ and is still rotationally invariant.  We may then set
$z=sf(s\bar s)$ once again and conclude that $\rho={\cal Q}(z)$
satisfies (\ref{m3.7}) and (\ref{m3.6}), for some coefficients
$\rho_n$ to be determined (up to a phase).  We define $${\cal
Z_A}=c_A\cdot \rho^{p_A}$$ so that the only equation to be solved is
$$[{\cal Z}_A, {\cal Z}_A^\dagger] = \sum_A |c_A|^2 [\rho^{p_A},
{\rho^\dagger}^{p_A}] = -\epsilon.$$ This can be rewritten in terms of
the coefficients $\alpha_n=|\rho_n|^2$ as
\be\label{m3.8}
\sum_A |c_A|^2 (\alpha_n\dots \alpha_{n+p_A-1}- \alpha_{n-1}\dots\alpha_{n-p_A})=\epsilon.
\ee
If one calls $$q=\max_A (p_A),$$ then equation (\ref{m3.8}) determines
$\alpha_{n+q-1}$ in terms of $\alpha_m$, for $m< n+q-1$.  On the other
hand, the coefficients $\alpha_0, \dots , \alpha_{q-2}$ are 
undetermined by (\ref{m3.8}).  We denote them by
$$\alpha_j=\xi_j.\hskip1cm (j=0,
\dots,q-2)$$ The $\alpha_n=\alpha_n(\xi)$ are then rational functions
of the $\xi_j$ and, 
as in the previous example, there should be a unique $\xi_j$ such that
$\alpha_n(\xi)\ge 0$ for all $n\ge 0$.

\subsection{General holomorphic curves}
\label{sec:example-general}

The examples that we have discussed so far can all be analyzed by
elementary methods.  On the other hand, the techniques used in these
special cases cannot easily be generalized.  First of all, we have
used the fact that $\Sigma={\bb C}$, allowing us to use the canonical
quantization of the plane which is well-known from elementary quantum
mechanics.  More complicated surfaces will have different underlying
spaces $\cal H$, and the examples given above do not suggest a natural
choice of $\cal H$ in the general case.  Moreover, even if we restrict
ourselves to the case $\Sigma = {\bb C}$, it is hard to find a general
solution of the problem.  The above
examples all rely on the rotational symmetry of the volume form $\mu$,
which allows us to conclude that $\rho$ is an operator satisfying $[N,
\rho] = \rho$.  This constraint restricts the form of $\rho$ almost completely.
In general the operator $\rho$ will have matrix elements between
eigenstates with arbitrary $N$ eigenvalues, and the operator-ordering
ambiguities will not allow us to explicitly determine the operator
$\rho$, or even to determine its existence.

The idea that allows us to solve the general case comes, on the other
hand, from the above examples if one changes point of view.  Let us go
back to the planar and parabolic examples.  In both cases the
underlying Hilbert space $\cal H$ was the same, but the quantization
of the analytic coordinate $z$ led to different operators, reflecting
the difference in the volume form $\mu$.  Let us suppose, on the other
hand, that the space $\cal H$, considered now just as a vector space
(therefore forgetting the inner product), does not change between the
two examples, and let us also suppose that the operator $\rho={\cal
Q}(z)$ is the same.  In both the planar and parabolic case we assume
that $\cal H$ is spanned by states $|v_n\rangle$, $n\ge 0$ (which are
proportional to the states $|n\rangle$) and we take
$$\rho |v_n\rangle = |v_{n+1}\rangle.$$ What changes between the
examples, reflecting the change in $\mu$, is the inner product on
$\cal H$ and therefore the definition of adjoint operator
$\rho^\dagger$.  We will assume that
$$\langle v_n|v_m \rangle=k_n \delta_{n,m}$$ so that $$\rho^\dagger
|v_n\rangle = {k_n\over k_{n-1}} |v_{n-1}\rangle.$$ (the absolute
normalization of the $k_n$'s is irrelevant, and we fix it by assuming
$k_0=1$).  We now have to solve for the coefficients $k_n$ by
imposing equation (\ref{m2.8}).  To make contact with the previous
discussion, we define $$\alpha_n={k_{n+1}\over k_n}.$$ Solving for the
$k_n$'s is equivalent to solving for the $\alpha_n$'s.

Let us look at the planar case.  Equation (\ref{m2.8}) is simply $[\rho,\rho^\dagger]=-\epsilon$ or,
in terms of the $\alpha_n$'s, $\alpha_n-\alpha_{n-1}=\epsilon$.  The recursive equation is solved by
$\alpha_n= (n+1)\epsilon$, or $k_n=n!\epsilon^n$.  The orthonormal states are then given by 
$|n\rangle =(n!\epsilon^n)^{-{1\over 2}} |v_n\rangle$, so that $\rho |n\rangle = \sqrt{\epsilon(n+1)}
|n+1\rangle$ as expected (recall $\rho=\sqrt\epsilon a^\dagger$ in the planar case).

The parabolic case is solved exactly as before, even though the interpretation is different.
Equation (\ref{m2.8}) reads $[\rho,\rho^\dagger]+[\rho^2, {\rho^\dagger}^2]=-\epsilon$ and it
implies again the recursive relation (\ref{m3.5}) for the coefficients $\alpha_n$.

In order to generalize the point of view described above we first have to decide how to choose,
given a surface $\Sigma$, a vector space ${\cal H}_\Sigma$ on which operators act.  A natural choice,
which is also suggested from coherent state quantization and geometric quantization (more of this in section \ref{f2}),
is to let 
$$
{\cal H}_\Sigma = \{{\rm holomorphic\ functions\ on\ } \Sigma\}.
$$
Note that we have not yet chosen an inner product on the space $\cal H$.
In the case $\Sigma = {\bb C}$, the states $|v_n\rangle$ correspond, up to an overall 
normalization, to the functions $z^n$.

Quantization of holomorphic functions, as we proposed before, will not
depend on the specific inner product chosen, and will be defined as
follows.  Let $|\phi\rangle$ be a state in $\cal H$ corresponding to
the holomorphic function $\phi$, and let $A$ be also holomorphic.  The
quantized operator ${\cal Q}(A)$ corresponding to $A$ will then be
defined by
$${\cal Q}(A) |\phi\rangle = |A\phi\rangle,$$ and it therefore acts by
pointwise multiplication of functions.  It is clear that operators
corresponding to holomorphic functions commute, and that equation
(\ref{m2.6}) is automatically satisfied.  Moreover, if the coordinate
functions $Z_A$ satisfy (\ref{m2.7}) then the quantized operators
${\cal Z}_A$ automatically satisfy (\ref{m2.9}).  In the specific
example given above we had $|v_n\rangle = |z^n\rangle$ and $\rho={\cal
Q}(z)$, so that $\rho|v_n\rangle = |v_{n+1}\rangle$, as we assumed
before.

Finally, as in the examples of this section, the crucial constraint
comes from equation (\ref{m2.8}).  We need to choose an inner product
on $\cal H$ in order to define adjoints of operators.  Since states
correspond to functions, it is natural to define an inner product via
integration over $\Sigma$.  In order to do so, we need a volume
$2$-form on $\Sigma$, which we will denote by $\Omega$ (Note that
$\Omega$ is quite different from the measure $\mu$; in particular,
$\mu$ does not have a finite integral when $\Sigma$ is noncompact).
We can then define the inner product between two elements
$\phi,\psi\in\cal H$ by
$$\langle\psi|\phi\rangle = \int_\Sigma \bar\psi \phi\,\Omega.$$ In
the planar case the correct choice of volume form is given by
$$\Omega=-{1\over 2\pi i \epsilon} e^{-z\bar z/\epsilon}\,dz\wedge
d\bar z.$$

The rest of the paper is devoted to the description of how $\Omega$
can be chosen for any functions $Z_A$ so that equation (\ref{m2.8}) is
satisfied.  In order to find such a description, we need to discuss in
more detail how to compute the adjoint of operators, and we will
therefore have to make a major mathematical digression devoted to the
study of Bergman integral kernels.

%%%%%%%%%%%%%%%%%%%%%%%%%%%%%%%%%%%%%%%%%%%%%%%%%%%%%%%%%%%%%%%%%%%%%%%%%
\section{Bergman Integral Kernels}
\label{bk0}
%%%%%%%%%%%%%%%%%%%%%%%%%%%%%%%%%%%%%%%%%%%%%%%%%%%%%%%%%%%%%%%%%%%%%%%%%

Let us suppose, in the language of the end of the last section, that
$\phi,\psi\in {\cal H}$, and that $A$ is a holomorphic function on
$\Sigma$.  We have proposed that ${\cal Q}(A)|\phi\rangle=
|A\phi\rangle$, and therefore we have that $\langle\psi|{\cal
Q}(A)|\phi\rangle=\int_\Sigma \bar\psi A\phi\,
\Omega$.  If we take the adjoint of the previous equation and use the fact that ${\cal Q}(A)^\dagger=
{\cal Q}(\bar A)$, we then obtain
$$\langle\phi|{\cal Q}(\bar A)|\psi\rangle=\int_\Sigma \bar\phi \bar
A\psi\,\Omega.$$ We might be tempted to deduce from the above equation
that ${\cal Q}(\bar A)|\psi\rangle= |\bar A\psi\rangle$, but this is
wrong since $\bar A\psi\not\in {\cal H}$.  On the other hand, if we
consider the larger space $\cal V$ of functions (not necessarily
analytic) on $\Sigma$ and view $\cal H$ as a subspace of $\cal V$,
then the above equation says that the orthogonal projection of the
state $|\bar A\psi\rangle$ onto $\cal H$ is equal to the state ${\cal
Q}(\bar A)|\psi\rangle$.  The orthogonal projection is called Bergman
projection, and this section is devoted to a detailed study of its
properties.  The Bergman integral kernel is defined in subsection
\ref{sec:bk-def}.  Subsection \ref{sec:geometry} contains a brief
review of the geometry of Riemann surfaces.  Some basic properties of
the Bergman integral kernel are discussed in \ref{sec:bk-properties}.
Subsection \ref{sec:bk-example} describes a particular example where
the Bergman projection operator can be explicitly computed; the form
of this projection operator suggests an ansatz for formulating the
operator in the general case, which is analyzed in subsection
\ref{sec:bk-general} and shown to agree with Bergman projection on a
general class of functions.

\subsection{Definition of Bergman integral kernels}
\label{sec:bk-def}

Consider a Riemann surface $\Sigma$ and fix on the surface, once and
for all, a real and non-vanishing $2$-form $\Omega$.  We  denote
by $\cal V$ the space of complex functions $f$ on $\Sigma$ which are
square integrable with respect to $\Omega$ - i.e.  such that
$\int_\Sigma |f|^2 \Omega < \infty$.  The vector space $\cal V$ has a
natural Hilbert space structure, where the inner product between two
elements $\phi,\psi\in\cal V$ is given by

$$
\langle\phi |\psi\rangle = \int_\Sigma \bar\phi \psi\ \Omega.
$$
The surface $\Sigma$ is endowed with a complex structure, and one is
therefore led, following Bergman \cite{bergman}, to consider the
subspace $\cal H \subset \cal V$ consisting of holomorphic functions
on $\Sigma$.  Moreover, since $\cal V$ is a Hilbert space, one can
study the orthogonal projection $\pi: \cal V \rightarrow \cal H$.
(The projection operator does depend on the choice of $2$-form
$\Omega$.  If we want to underline this dependence, we will use the
more cumbersome notation $\pi_\Omega$).  In what follows we wish to
give an integral representation of the projection operator and to
study its properties.  To this end we fix an orthonormal basis for
$\cal H$ given by holomorphic functions $f_n$ on $\Sigma$ satisfying
$\int_\Sigma \bar f_n f_m\ \Omega = \delta_{n,m}$.  
(Note that this basis is unrelated to the basis $| v_n \rangle$
discussed in the previous section.)
The orthogonal
projection of any element $\phi\in\cal V$ is given by $\pi(\phi)=
\sum_n f_n \langle f_n | \phi \rangle$.  We may then introduce a
kernel function $K$, called the Bergman integral kernel, defined by

\be\label{bk1.0}
K(z,w)=\sum_n f_n(z)\bar f_n(w).
\ee
The function $K$ does not depend on the specific choice of basis $f_n$, since different choices
are related by unitary transformations.  Moreover it gives an integral representation 
of the action of the projection operator $\pi$.  In fact, if $\phi$ is an element of $\cal V$,
one has that

$$\pi(\phi)(z) = \int_\Sigma K(z,w) \phi(w)\ \Omega(w).
$$

\subsection{Geometry of Riemann surfaces}
\label{sec:geometry}

In order to gain a deeper understanding of the action of the Bergman
projection $\pi$, we need to be able to discuss in more detail the
geometry of the underlying Riemann surface $\Sigma$.  To this end we
could very well use the standard notation of differential and
Riemannian geometry.  On the other hand, the fact that the manifold
$\Sigma$ has a complex structure and is of complex dimension one,
greatly simplifies the geometry and the standard notation is very
cumbersome in this specific case.  Therefore we use this subsection to
introduce some specific conventions and notations which will simplify
the manipulations and, hopefully, clarify the underlying geometric
concepts.

We first focus our attention on tensors, which are classified
according to their conformal weight (or dimension).  A tensor $T$ has
conformal weight $(a,\bar a)$ if, under coordinate transformations,
the expression

$$
T\ dz^a d{\bar z}^{\bar a}
$$
is invariant.  We can use various operations to construct new tensor
fields starting from old ones.  Some of these manipulations are
standard, like tensor addition and multiplication\footnote{Recall
that addition is defined for tensors of the same weight and
multiplication is defined for tensors of any weight.  Under
multiplication, the conformal dimension of the resulting tensor is the
sum of the dimensions of the original tensors.}.  Some operations, on the
other hand, are specific to the case of a complex variety of dimension
one, and will be used repeatedly in the rest of the paper.  Let us
start, for example, with a non-vanishing tensor field $T$ of conformal
weight $(a,\bar a)$.  Since tensor fields are sections of line
bundles, we can consider the inverse field

$$1/T
$$
of dimension $(-a,-\bar a)$.  As a second example we may start with the same tensor field $T$ and construct
a new field 
\be\label{bk1.-1}
\partial\bar\partial\ \ln\ T
\ee
of dimension $(1,1)$.  Some explanation is necessary in this case.
First of all we must be able to consistently choose a branch of the
logarithm.  This is possible for example if $a=\bar a$ and if $Arg(T)$
is constant (note that this is a well defined notion because, under
change of coordinates $z\rightarrow w$, we have that $T\rightarrow T
|\partial w/\partial z|^{2a}$, so that $Arg(T)$ is invariant).
Secondly we must show that the expression in equation (\ref{bk1.-1})
does define a tensor of the correct dimension.  Under a coordinate
transformation $z\rightarrow w$ we have that $\ln(T)\rightarrow
\ln(T)+a\ \ln(\partial w/\partial z) + \bar a\ \ln(\bar\partial \bar w/\bar \partial \bar z)$.  
The second and third term in the transformation law are respectivelly analytic and antianalytic, 
and are therefore annihilated by the operator $\partial\bar\partial$.  Therefore $\partial_z\bar\partial_{\bar z}
\ \ln(T)\rightarrow \partial_w\bar\partial_{\bar w}\ \ln(T)\ |\partial w/\partial z|^2$, as we wanted to show.

Up to this point we have used the $2$-form $\Omega$ to define the integration measure on the surface $\Sigma$.
It will be very useful later to consider $\Omega$ as the volume form of an underlying riemannian metric 
$g$.  Let us be more specific.  If $z=x+iy$ is a local analytic coordinate, we can write

$$
\Omega = i\ C dz \wedge d\bar z,
$$
where $C$ is a real and positive $(1,1)$ tensor.  We will choose $g$ so that 

$$
\Omega=\sqrt{\det\ g_{ab}}\ dx\wedge dy.
$$
We clearly have some freedom in our choice of the metric.  If we impose the additional restriction that $g$ be hermitian 
-- i.e.  that

$$
g_{zz}=g_{\bar z \bar z}=0,
$$
then the only non-vanishing element of the metric is

$$
g_{z\bar z}=C.
$$
The standard riemannian connection is in this case very simple.  The only
non-vanishing coefficients are given by

\bea
\Gamma =& \Gamma ^{z}_{zz} = \partial\ \ln\ C \nonumber\\
\bar\Gamma =& \Gamma^{\bar z}_{\bar z\bar z} = \bar\partial\ \ln\ C.\nonumber
\eea
The covariant derivatives of a tensor $T$ of weight $(a,\bar a)$ can then be written in terms of
the connection as

\bea
\nabla T &=(\partial - a\Gamma)T  \nonumber\\
\bar\nabla T &=(\bar\partial-\bar a\bar\Gamma)T.   \nonumber
\eea
We conclude this quick tour of Riemannian geometry by considering the
curvature tensor, which measures the lack of commutativity of
covariant derivatives.  It is a simple computation to check that

\be\label{bk1.1}
[\nabla,\bar\nabla] T = (a-\bar a) R\,T,
\ee
where $R$ is the $(1,1)$ curvature tensor, given by

$$
R=\partial\bar\Gamma=\bar\partial\Gamma=\partial\bar\partial\ \ln\ C.
$$
As a final remark we note that the tensor in equation (\ref{bk1.-1}) can be rewritten in terms of covariant
derivatives as

\bea
\partial\bar\partial\ \ln\ T =&\nabla{1\over T}\bar\nabla T + \bar a R  \nonumber\\
                            =&\bar\nabla{1\over T}\nabla T + a R.     \nonumber
\eea
Note that we will use the  convention that covariant derivatives act on  everything
on the right, unless explicitly indicated.

\subsection{Properties of Bergman integral kernels}
\label{sec:bk-properties}

We now have the language to discuss some of the basic properties of
the Bergman projection.  The first property follows essentially from
the definition.  If $\phi$ is an element of $\cal V$, then $\pi(\phi)$
is in $\cal H$, and is therefore analytic.  Moreover, if $\phi$ itself
is analytic, then $\pi(\phi)=\phi$.  The second property of the
projection requires more work.  We start by observing that $K(z,w)$ is
analytic in $z$ and antianalytic in $w$, as can be readily seen from
the defining equation (\ref{bk1.0}).  We then let $X$ be a $(-1,0)$
vector field and we consider the scalar field $\nabla X \in \cal V$.
Using the integral representation of $\pi$ we can compute, recalling
that $\nabla_w K(z,w)=0$,

\bea
\pi(\nabla X)(z) =& \int_\Sigma K(z,w) 
\nabla_w X(w)\ \Omega(w)=     \nonumber\\
                 =& -\int_\Sigma X(w)\nabla_w K(z,w) \ \Omega(w)=0,
                 \nonumber
\eea
where we assume, as we will from now on, that we can neglect boundary
terms when we integrate by parts (this is  true in all the cases of
interest in this paper).  Note that the above manipulation is possible
since $\Omega$ is the volume form of the underlying metric, and
therefore integration by parts can be performed, provided we replace partial
derivatives with covariant derivatives.  We have thus shown that,
generically,

$$
\pi(\nabla X)=0.
$$

We may combine the two properties described above as follows.  Let $X$ and $\phi$ be, respectively, a holomorphic $(-1,0)$
vector field and a holomorphic function.  Then the function $X\nabla\phi$ is itself holomorphic, and therefore
$\pi(X\nabla\phi)=X\nabla\phi$.  Moreover, as we have shown above, $\pi(\nabla X\phi)=0$.  We can use the fact that
$\nabla X\phi = X\nabla\phi
+ \phi\nabla X$ and the fact that $\pi$ is a linear map to conclude that $\pi(\phi\nabla X)=-X\nabla\phi$.
Using the same reasoning inductively we can show that
\be
\pi(\phi \nabla X_1\nabla X_2\dots \nabla X_n)=(-1)^n X_n\nabla
X_{n-1}\nabla \dots X_1\nabla \phi, 
\label{eq:general-pi}
\ee
where the $X_i$ are holomorphic $(-1,0)$ vector fields, and $\phi$ is a holomorphic function.  

We have already remarked that the projection operator depends
implicitly on the underlying $2$-form $\Omega$.  Changes in $\Omega$
are reflected non-trivially in changes in $\pi$.  In general this
relationship is quite complicated.
In one specific
case,  however, we can explicitly relate the projection
operators corresponding to different choices of $\Omega$.  Let $\chi$
be a holomorphic function and consider the following transformation

$$
\Omega\rightarrow e^{\chi+\bar\chi}\ \Omega.
$$
Any orthonormal basis $f_n$ of $\cal H$
undergoes the corresponding transformation $f_n\rightarrow f_n e^{-\chi}$ and, therefore, the integral kernel is modified
as follows

$$
K(z,w)\rightarrow e^{-\chi(z)}K(z,w)e^{-\bar\chi(w)}.
$$
Using the integral representation of the Bergman projection, it is then the work of a minute to show that, for any function
$\phi$ 

\be\label{bk1.2}
\pi_{\Omega e^{\chi+\bar\chi}}(\phi)= e^{-\chi} \pi_\Omega (\phi e^{\chi}).
\ee

%%%%%%%%%%%%%%%%%%%%%%%%%%%%%%%%%%%%%%%%%%%%%%%%%%%%%%%%%%%%%%%%%%%%%%%%%
\subsection{An example of Bergman projection}
\label{sec:bk-example}
%%%%%%%%%%%%%%%%%%%%%%%%%%%%%%%%%%%%%%%%%%%%%%%%%%%%%%%%%%%%%%%%%%%%%%%%%

We have completed an informal discussion of the Bergman projection and
of its basic properties.  We now focus our attention once more on the
integral representation of the projection $\pi$.  Given a function
$\phi\in\cal V$, the value $\pi(\phi)(z)$ will depend generically on
the values $\phi(w)$ for arbitrary $w$, and in this sense the operator
$\pi$ is non-local.  On the other hand, as we will show in detail in
this subsection and the following subsection,
there is a useful and explicit expansion for $\pi(\phi)(z)$ in terms
of the values of $\phi$ and of its covariant derivatives, all evaluated
at the same point $z$.  The expansion is schematically of the form
\be\label{bk1.10}
\pi(\phi)(z)\sim \sum_{n=0}^\infty {1\over R^n} \nabla^n\bar\nabla^n \phi (z).
\ee
The interesting feature is that, for surfaces with very large
curvature, the first terms dominate the sum and the operator becomes
essentially local.  We will see that this limit is very natural in the
context of geometric quantization of holomorphic surfaces, and
corresponds to the limit of zero Planck constant (the classical limit).

In this subsection we describe in detail a simple example where the
Bergman projection operator can be computed explicitly.  We
obtain an expression for the action of this projection operator of the
form just described, and this will motivate the ansatz for the general
form that is examined in  subsection
\ref{sec:bk-general}.

We take $\Sigma$ to be the complex plane ${\bb C}$ and $$\Omega=-{1\over
2\pi i \epsilon} e^{-z\bar z/ \epsilon} dz\wedge d\bar z,$$ where
$\epsilon$ is a positive constant.  A natural orthonormal basis for
the space $\cal H$ is given by the functions

$$
f_n(z)= c_n z^n,
$$
where 
$$
|c_n|^{-2} = \int_\Sigma  \bar z^n z^n\ \Omega= \epsilon^{n} n!.
$$ 
The Bergman integral kernel can be explicitly computed
$$
K(z,w)=\sum_{n=0}^\infty {1\over \epsilon^{n}n!} z^n\bar w^n= e^{z\bar w /\epsilon},
$$and therefore the integral representation of the projection operator is explicitly given by

\be\label{bk2.1}
\pi(\phi)(z) = -{1\over\epsilon}{1\over 2\pi i}\int_\Sigma e^{(z-w)\bar w/\epsilon}\phi(w) dw\wedge d\bar w,
\ee
where $\phi\in\cal V$.  In order to analyze the above integral we will first of all introduce some auxiliary 
functions

\be\label{bk2.2}
G_n={\epsilon^n\over(w-z)^n}e^{(z-w)\bar w/\epsilon}
\ee
for $n\geq 0$.  Note that equation (\ref{bk2.1}) can be rewritten as 
$$
\pi(\phi)(z)= -{1\over\epsilon}{1\over 2\pi i} \int_\Sigma G_0 \phi\ dw\wedge d\bar w.
$$
Moreover
\bea
G_0 \phi\ dw\wedge d\bar w &=& d_w(G_1\phi\ dw)+G_1 \bar\partial_w \phi\ dw\wedge d\bar w   \nonumber\\
   &\vdots &      \nonumber\\
G_n \bar\partial_w^{(n)} \phi\ dw\wedge d\bar w &=& d_w(G_{n+1}\bar\partial_w^{(n)} \phi\ dw)+
G_{n+1} \bar\partial_w^{(n+1)} \phi\ dw\wedge d\bar w      \nonumber
\eea
so that
$$
G_0\phi\ dw\wedge d\bar w = d_w(\sum_{n=0}^\infty G_{n+1} \bar\partial_w^{(n)}\phi\ dw).
$$
All of the functions $G_n$, for $n\geq 1$, are singular at $w=z$.  This suggests that we should replace the 
integration region $\Sigma$ with the region $\Sigma_\delta$, obtained from the full complex plane by deleting a disk
of radius $\delta$ around $z$.  It will be then understood in the sequel that we are considering the limit
$\delta\rightarrow 0$.  We may then use Stokes theorem and write

\bea
\pi(\phi)(z) &=& -{1\over\epsilon}{1\over 2\pi i}\int_{\Sigma_\delta} 
d_w(\sum_{n=0}^\infty G_{n+1} \bar\partial_w^{(n)}\phi\ dw) \nonumber\\
&=& \sum_{n=0}^\infty {1\over\epsilon} \int_{\Gamma_\delta} {dw\over 2\pi i}  G_{n+1} \bar\partial_w^{(n)}\phi,\nonumber
\eea
where $\Gamma_\delta =-\partial\Sigma_\delta$ is the circle of radius $\delta\rightarrow 0$ around $z$, with counterclockwise
orientation.  We first recall that for any function, not necessarily analytic,
$$
\lim_{\delta\rightarrow 0} \int_{\Gamma_\delta} {dw\over 2\pi i} {A(w)\over (w-z)^{n+1}} = {1\over n!}\partial^{(n)}A(z).
$$
Therefore, using equation (\ref{bk2.2}) for $G_n$, we obtain

\bea
\pi(\phi)(z) &=& \sum_{n=0}^\infty {\epsilon^n\over n!} \partial_w^{(n)} \left( e^{(z-w)\bar w/\epsilon}\bar\partial_w^{(n)}
\phi(w) \right) \big|_{w=z} \nonumber\\
&=& \sum_{n=0}^\infty {\epsilon^n\over n!} \sum_{p=0}^n \left(\begin{array}{c} n\\p\end{array}\right) \left(-{\bar z\over
\epsilon}\right)^p\partial^{n-p}\bar\partial^n\phi(z)\nonumber
\eea
and finally

\be\label{bk2.3}
\pi(\phi) = \sum_{n=0}^\infty {\epsilon^n\over n!} (\partial-{\bar z\over\epsilon})^n\bar\partial^n\phi.
\ee
Up to this point we have used a specific coordinate system and therefore the geometric nature of the above expression
is not transparent.  Let us therefore rewrite equation (\ref{bk2.3}) in a coordinate invariant way.  First note that the
geometric data for the surface is given by

\bea
C &=& {1\over 2\pi\epsilon} e^{-z\bar z/\epsilon}\nonumber\\
\Gamma &=& -{\bar z\over \epsilon}\hskip1cm \bar\Gamma = -{z\over\epsilon}\nonumber\\
R &=& -{1\over\epsilon}.\nonumber
\eea
We first of all note that the expansion coefficient $\epsilon$ is
inversely related to curvature, and we start to see the first evidence
for the claim (\ref{bk1.10}).  Moreover the holomorphic derivatives
$(\partial-\bar z/\epsilon)$ and antiholomorphic derivatives
$\bar\partial$ can be replaced with covariant derivatives $\nabla$ and
$\bar\nabla$ as long as they are acting on tensors of holomorphic
dimension $-1$ and antiholomorphic dimension $0$ respectively.  This
can be easily done by writing the $n$-th term of the sum in equation
(\ref{bk2.3}) as

\be\label{bk2.4}
{1\over n!}(-1)^n \left(\nabla {1\over R}\right)^n R^n \left({1\over R}\bar\nabla\right)^n \phi
\ee
First note that the above expression has the right power of curvature
to give a total contribution of $\epsilon^n$ (the minus signs reflect
the fact that $R=-1/\epsilon$).  Starting from the right of equation
(\ref{bk2.4}) we can also see that the covariant derivatives act on
tensors of the correct dimension.  The first antiholomorphic
derivative $\bar\nabla$ acts on a $(0,0)$ tensor, thus giving a
$(0,1)$ tensor.  After dividing by $R$, we get a $(-1,0)$ tensor.
Repeating this process $n$ times we see that $({1\over R}
\bar\nabla)^n\phi$ has weight $(-n,0)$ and so
$$
R^n \left({1\over R}\bar\nabla\right)^n \phi
$$
has conformal dimension $(0,n)$.  Division by $R$ moves the dimension
to $(-1,n-1)$ and multiplication by $\nabla$ moves it to $(0,n-1)$.
Continuing this process we see that $\nabla$ always acts on tensors of
holomorphic dimension $-1$ and $\bar\nabla$ on tensors of
antiholomorphic dimension $0$.  Replacing $\nabla\rightarrow \partial
+ \Gamma$ and $\bar\nabla\rightarrow\bar\partial$ in equation
(\ref{bk2.4}) and using $R=-1/\epsilon$ we then recover the $n$-th
term in the sum (\ref{bk2.3}).  We thus conclude that

\be\label{bk2.5}
\pi(\phi)=\sum_{n=0}^\infty {1\over n!}(-1)^n \left(\nabla {1\over R}\right)^n R^n 
\left({1\over R}\bar\nabla\right)^n \phi.
\ee
Recall that we are using a convention in which the covariant
derivatives act on everything to their right, even when they are
inside a parentheses.
At this point we could analyze the above expression further and check
that it satisfies the properties of the Bergman projection described
in the previous section.  On the other hand, this will follow as a
special case of the general discussion of the next section.  We will
be therefore content to use equation (\ref{bk2.5}) as a partial
motivation for the general ansatz that will be discussed next.

%%%%%%%%%%%%%%%%%%%%%%%%%%%%%%%%%%%%%%%%%%%%%%%%%%%%%%%%%%%%%%%%%%%%%%%%%
\subsection{The general Bergman projection operator}
\label{sec:bk-general}
%%%%%%%%%%%%%%%%%%%%%%%%%%%%%%%%%%%%%%%%%%%%%%%%%%%%%%%%%%%%%%%%%%%%%%%%%

In this section we shall argue that the general expansion for the
action of the Bergman projection on a generic function $\phi\in\cal V$
is given by
(again, recall that covariant
derivatives act on everything to their right, even when they are
inside a parentheses.)

\be\label{bk3.1}
\pi(\phi)= \sum_{n=0}^\infty (-1)^n (\nabla {1\over P_1})(\nabla {1\over P_2})\dots (\nabla {1\over P_n}) P_1\dots P_n
({1\over P_n}\bar\nabla)\dots({1\over P_2}\bar\nabla)({1\over P_1}\bar\nabla)\phi,
\ee
where the $P_n$'s are $(1,1)$ tensors related to the curvature tensor
$R$ and its derivatives.  
In particular

$$
P_1=R,
$$ 
and the $P_n$'s, for $n>1$, satisfy the recursion relation

\be\label{bk3.2}
P_n=P_1+P_{n-1}+\sum_{j=1}^{n-1}\partial \bar \partial\ \ln\ |P_j|
=P_1+P_{n-1}+\sum_{j=1}^{n-1}\partial \bar \partial\ \ln\ P_j.
\ee
We will show in particular that the expansion (\ref{bk3.1}) satisfies all of the properties of the projection
operator described in section \ref{sec:bk-properties}.

Before giving the formal proof let us show that expression
(\ref{bk3.1}) reduces to equation (\ref{bk2.5}) if we are considering
the special case of the last section.  We recall that, in the
canonical coordinate system that we chose in section \ref{sec:bk-example}, the
curvature tensor was $R=-1/\epsilon$, and was therefore independent of
$z$.  Using the recursive equation (\ref{bk3.2}) one can then show
inductively that all the $P_n$'s are independent of $z$ and therefore
the terms of the form $\partial\bar\partial\ \ln\ P_j$ drop from the
recursion relation.  It is then easy to show that $P_n= n R$,
therefore recovering equation (\ref{bk2.5}).

Now back to the main proof.  First we note that, if $\phi$ is
holomorphic, all the terms with $n\ge1$ in the expansion (\ref{bk3.1})
vanish, and therefore $\pi(\phi)=\phi$ as expected.  We will now show
that, for a generic $\phi$, $\pi(\phi)$ is analytic - i.e.  that
$\bar\nabla\pi(\phi)=0$.  Let $\pi_n$ be the $n$-th term in the sum
(\ref{bk3.1}), so that $\pi(\phi)=\sum_{n=0}^\infty\pi_n.$ We write
the expression for $\bar\nabla \pi_n$ as a sum

$$
\bar\nabla\pi_n =A_n+B_n,
$$
where

\bea
A_n &=& (-1)^n [\bar\nabla,(\nabla {1\over P_1})  \dots (\nabla {1\over P_n}) P_1\dots P_n]
({1\over P_n})\bar\nabla \dots ({1\over P_1}\bar\nabla)\phi\nonumber\\
B_n &=& (-1)^n (\nabla {1\over P_1})\dots (\nabla {1\over P_n}) P_1\dots P_n P_{n+1}
({1\over P_{n+1}}\bar\nabla)\dots({1\over P_1}\bar\nabla)\phi.  \nonumber
\eea
Clearly $A_0=0$.  Moreover we will show that $A_n+B_{n-1}=0$ for $n\geq 1$.  This will complete the proof,
because $\bar\nabla\pi(\phi)=\sum_{n=0}^\infty \bar\nabla\pi_n=\sum_{n=0}^\infty (A_n+B_n)=
A_0+\sum_{n=1}^\infty (A_n+B_{n-1})=0$.  We shall use induction to show that $A_n+B_{n-1}=0$.  It is 
convenient to this end to write the expression for $A_n$ and $B_n$ as 

\bea
A_n &=& (-1)^n \alpha_n ({1\over P_n}\bar\nabla)\dots({1\over P_1}\bar\nabla)\phi \nonumber\\
B_n &=& (-1)^n \beta_n ({1\over P_{n+1}}\bar\nabla)\dots({1\over P_1}\bar\nabla)\phi, \nonumber
\eea
where $\alpha_n$ and $\beta_n$ are operators acting on tensors of weight $(-n,0)$ and $(-n-1,0)$ respectively, and
are given by

\bea
\alpha_n &=& [\bar\nabla,(\nabla {1\over P_1})  \dots (\nabla {1\over P_n}) P_1\dots P_n]\nonumber\\
\beta_n &=& (\nabla {1\over P_1})\dots (\nabla {1\over P_n}) P_1\dots P_n P_{n+1}.\nonumber
\eea
It will then be sufficient to show that $\alpha_n=\beta_{n-1}$ when acting on tensors of dimension $(-n,0)$.
The case $n=1$ is a simple application of equation (\ref{bk1.1}) for the commutator of covariant derivatives acting
on $(-1,0)$ vector fields

$$
\alpha_1=[\bar\nabla,\nabla]=R=P_1=\beta_0.
$$
The induction step, on the other hand, is proved as follows.  Rewrite
the expression for $\alpha_n$ as

\bea\label{bk3.3}
\alpha_n &=& [\bar\nabla,(\nabla {1\over P_1})  \dots (\nabla {1\over P_{n-1}}) P_1\dots P_{n-1}{1\over P_1\dots P_{n-1}}
(\nabla{1\over P_n})P_1\dots P_n]\nonumber\\
&=& \alpha_{n-1}({1\over P_1\dots P_{n-1}} \nabla P_1\dots P_{n-1})+\\
&& + (\nabla {1\over P_1})  \dots (\nabla {1\over P_{n-1}}) P_1\dots P_{n-1} [\bar\nabla, 
{1\over P_1\dots P_{n-1}} \nabla P_1\dots P_{n-1}].\nonumber
\eea
Recall that the above operator acts on $(-n,0)$ tensors.  The commutator in the second term of the above expression
for $\alpha_n$, which we will denote by $O$, can be rewritten as

\bea
O &=& [\bar\nabla, {1\over P_1\dots P_{n-1}} \nabla P_1\dots P_{n-1}]\nonumber\\
 &=& [\bar\nabla, {1\over P_1\dots P_{n-1}} [\nabla, P_1\dots P_{n-1}]]+[\bar\nabla,\nabla]\nonumber\\
&=& T+nR,\nonumber
\eea
where the tensor $T$ is given by 

$$
T=\bar\nabla {1\over P_1\dots P_{n-1}} \nabla P_1\dots P_{n-1}= \partial\bar\partial\ \ln\ P_1\dots P_{n-1}-(n-1)R.
$$
Then
 
$$
O = R+\sum_{j=1}^{n-1} \partial\bar\partial\ \ln\ P_j=P_n-P_{n-1}.
$$
We can then combine the above result with the inductive hypothesis
$\alpha_{n-1}=\beta_{n-2}$ and rewrite equation (\ref{bk3.3}) as

\bea
\alpha_n &=& \beta_{n-2}{1\over P_1\dots P_{n-1}}\nabla P_1\dots P_{n-1} +
(\nabla {1\over P_1})  \dots (\nabla {1\over P_{n-1}}) P_1\dots P_{n-1} O =\nonumber\\
&=& (\nabla {1\over P_1})  \dots (\nabla {1\over P_{n-2}}) \nabla P_1\dots P_{n-1} +
(\nabla {1\over P_1})  \dots (\nabla {1\over P_{n-1}}) P_1\dots P_{n-1} P_n \nonumber\\
&& -(\nabla {1\over P_1})  \dots (\nabla {1\over P_{n-2}})\nabla P_1\dots P_{n-1} = \beta_{n-1},\nonumber
\eea
thus proving the inductive step.

The next property of the projection operator that we want to prove is that $\pi(\nabla X)=0$ when $X$ is a $(-1,0)$
vector field.  The proof is similar to the one just given, and we shall be brief.  Suppose that $\phi=\nabla X$ in
equation (\ref{bk3.1}).  The expression for $\pi_n$ can be written, using the same philosophy, 
as the sum of two terms $$\pi_n = A_n + B_n,$$
where now

\bea
A_n &=& (-1)^{n+1} (\nabla {1\over P_1})\dots (\nabla {1\over P_n}) \alpha_n X   \nonumber \\
B_n &=& (-1)^n (\nabla {1\over P_1})\dots (\nabla {1\over P_{n+1}}) \beta_n X.  \nonumber
\eea
The operators $\alpha_n$ and $\beta_n$ are now given by
\bea
\alpha_n &=& [\nabla, P_1\dots P_n ({1\over P_n}\bar\nabla)\dots({1\over P_1}\bar\nabla)]            \nonumber\\
\beta_n &=& P_1\dots P_{n+1}({1\over P_n}\bar\nabla)\dots({1\over P_1}\bar\nabla)      \nonumber
\eea
and both act on $(-1,0)$ fields.  Again $A_0 = 0$ and the proof rests on the fact that $A_n+B_{n-1}=0$.  In fact we
will prove, like above, that $\alpha_n+\beta_{n-1}=0$.  Clearly $\alpha_1=[\nabla,\bar\nabla]=-R=-\beta_0$.  Moreover

\bea
\alpha_n &=& P_1\dots P_{n-1} \bar\nabla {1\over P_1\dots P_{n-1}} [\nabla, P_1\dots P_{n-1}
({1\over P_{n-1}}\bar\nabla)\dots({1\over P_1}\bar\nabla)] +   \nonumber \\
&& + [\nabla, P_1\dots P_{n-1} \bar\nabla {1\over P_1\dots P_{n-1}}] P_1\dots P_{n-1} 
({1\over P_{n-1}}\bar\nabla)\dots({1\over P_1}\bar\nabla) \nonumber \\
&=& P_1\dots P_{n-1} \bar\nabla {1\over P_1\dots P_{n-1}} \alpha_{n-1}+ O P_1\dots P_{n-1} 
({1\over P_{n-1}}\bar\nabla)\dots({1\over P_1}\bar\nabla).  \nonumber
\eea
As before we can show that, in this case, $O=P_{n-1}-P_n$.  Then

\bea
\alpha_n &=& -  P_1\dots P_{n-1} \bar\nabla ({1\over P_{n-2}}\bar\nabla)\dots({1\over P_1}\bar\nabla)+
P_1\dots P_{n-1} \bar\nabla ({1\over P_{n-2}}\bar\nabla)\dots({1\over P_1}\bar\nabla)   \nonumber\\
&&-P_1\dots P_{n} ({1\over P_{n-1}}\bar\nabla)\dots({1\over P_1}\bar\nabla) = - \beta_{n-1},    \nonumber
\eea
as was to be shown.  

We can use the fact that $\alpha_n+\beta_{n-1} = 0$ to prove an
interesting corollary.  Suppose that $X_1,\dots, X_n$ are holomorphic
$(-1,0)$ vector fields and that $\phi$ is a holomorphic function.  We
will prove that
\be\label{bk3.30}
({1\over P_n}\bar\nabla)\dots({1\over P_1}\bar\nabla)\,\phi\nabla X_1\dots \nabla X_n = \phi\,X_1\dots X_n.
\ee
Before we prove this fact, let us note the importance of equation
(\ref{bk3.30}).  We recall from section \ref{sec:bk-properties} that
$\pi(\phi\nabla X_1\dots \nabla X_n)=(-1)^n X_n\nabla \dots
X_1\nabla\phi$.  The left hand side of this equality can be expanded
using equation (\ref{bk3.1}), and it is given in general by an
infinite sum.  On the other hand, in this particular case, equation
(\ref{bk3.30}) implies that the sum in (\ref{bk3.1}) stops after $n$
terms, since the $(n+1)$-th term will contain an antiholomorphic
derivative $\bar\nabla$ acting on the left-hand side of equation
(\ref{bk3.30}).  Since the right-hand side is manifestly holomorphic,
the $(n+1)$-th term vanishes (and similarly all the $m$-th terms, with
$m>n$).

Now to the inductive proof.  First we note that we can take $\phi=1$ with no loss of generality, since $\phi$ commutes
with $\bar\nabla$.  We multiply equation (\ref{bk3.30}) by $P_1\dots P_n$, and we write, using the inductive hypothesis,
\bea
&& P_1\dots P_n({1\over P_n}\bar\nabla)\dots({1\over
P_1}\bar\nabla)\nabla X_1\dots \nabla X_n = \nonumber\\ && = \nabla
P_1\dots P_n({1\over P_n}\bar\nabla)\dots({1\over P_1}\bar\nabla)
X_1\nabla\dots \nabla X_n -
\alpha_n X_1\nabla\dots \nabla X_n = \nonumber\\
&& = \nabla P_1\dots P_{n-1} X_1 \bar\nabla X_2\dots X_n + \beta_{n-1}X_1\nabla\dots \nabla X_n =\nonumber\\
&& = P_1\dots P_n ({1\over P_{n-1}}\bar\nabla)\dots({1\over P_1}\bar\nabla) X_1\nabla\dots \nabla X_n=\nonumber\\
&& = P_1\dots P_n X_1\dots X_n,\nonumber
\eea
as was to be shown.

The last property of $\pi$ that we would like to check is its behavior
under a transformation $\Omega\rightarrow e^{\chi+\bar\chi}\Omega$,
which is described by equation (\ref{bk1.2}).  Under such a
transformation we have that $\ln\ C\rightarrow \chi+\bar\chi+\ln\ C$,
and therefore $R$ is invariant.  Moreover, since the recursion
relation (\ref{bk3.2}) does not involve the metric, all of the tensors
$P_n$ are invariant.  On the other hand the connection does change.
In particular $\Gamma \rightarrow \Gamma+\partial\chi$.  The expansion
in equation (\ref{bk3.1}) does not contain $\bar\Gamma$, since all of
the antiholomorphic derivatives act on tensors of antiholomorphic
dimension $0$.  On the other hand the holomorphic derivatives act on
tensors of holomorphic dimension $-1$, and can therefore be replaced
with $\partial + \Gamma$.  If we act with the expansion (\ref{bk3.1})
on the function $e^\chi \phi$, we may consider $e^\chi$ as an operator
and move it to the left.  It commutes with everything exept with the
holomorphic covariant derivatives and in this case we pick up a
commutator term $[\nabla,e^\chi]=\partial \chi e^\chi$.  This extra
term reflects the change in the connection coefficient $\Gamma$ given
above.  Therefore the net effect of moving the operator $e^\chi$ all
the way to the left is that $\nabla_\Omega \rightarrow\nabla_\Omega +
\partial\chi= \nabla_{e^{\chi+\bar\chi}\Omega}$.  If we multiply the
whole expansion by $e^{-\chi}$, we then obtain equation
(\ref{bk1.2}).

We have thus shown that the projection operator defined in
(\ref{bk3.1}) has all the properties we expect of the Bergman
projection operator, and that in particular the projection operators
agree on a very general class of functions on $\Sigma$, namely those
whose projections are described in (\ref{eq:general-pi}).  This does not
give a completely mathematically rigorous proof that the expression
(\ref{bk3.1}) correctly describes the action of Bergman projection on
an arbitrary function $\phi$.  A complete proof might follow, for
example, by showing that the set of functions on which the projection
operators agree is dense in the requisite function space.  For the
purposes of this paper, however, we can treat (\ref{bk3.1}) as the
definition of Bergman projection on a general function, and we can now
use this operator to construct the general matrix representation of
any holomorphic curve.

%%%%%%%%%%%%%%%%%%%%%%%%%%%%%%%%%%%%%%%%%%%%%%%%%%%%%%%%%%%%%%%%%%%%%%%%%
\section{Representation of General Holomorphic Curves}
\label{f1}
%%%%%%%%%%%%%%%%%%%%%%%%%%%%%%%%%%%%%%%%%%%%%%%%%%%%%%%%%%%%%%%%%%%%%%%%%

We have finally concluded our long digression on the Bergman integral
kernel and we can go back to the analysis of the original problem of
matrix representations of holomorphic curves.

We first define the quantization operator $\cal Q$ precisely.  Let
$\phi\in \cal H$ and let $A$ be a generic function on $\Sigma$.  The
operator ${\cal Q}(A)$ corresponding to $A$ is defined by
$$ {\cal Q}(A) |\phi\rangle = \pi |A\phi\rangle.$$ In words, we first multiply $\phi$ by the function $A$
and then we extract the holomorphic part using the Bergman projection.  We note a few simple properties of 
the definition.  First of all, the above definition is consistent with the one given at the end of 
section \ref{m3} when $A$ is holomorphic, since in this case $\pi|A\phi\rangle = |A\phi\rangle$.
We also note that, if $\phi,\psi\in\cal H$, then $$\langle \phi | {\cal Q}(A) | \psi\rangle =  
\langle \phi | A  \psi\rangle = \int_\Sigma \bar\phi A \psi\,\Omega,$$ since $\langle \phi|\pi = \langle\phi|$.
Moreover, under complex conjugation, we have that 
\bea
\langle \phi | {\cal Q}^\dagger(A) | \psi\rangle &=& \langle \psi | {\cal Q}(A) | \phi\rangle^*=
\int_\Sigma \bar\phi \bar A \psi\,\Omega\nonumber\\
&=&\langle \phi | {\cal Q}(\bar A) | \psi\rangle,\nonumber
\eea
so that we have $${\cal Q}(\bar A)= {\cal Q}^\dagger(A).$$ Finally it
is clear from the definition that ${\cal Q}(1) = {\rm Id}_{\cal H}$.

We are now in a position to analyze equation (\ref{m2.8}) in its full
generality.  We will find it convenient to consider the matrix element
of equation (\ref{m2.8}) between two states $\phi,\psi\in \cal H$, and
we  therefore study the equation
\be\label{f1.3}
\langle \phi |\, [{\cal Z}_A,{\cal Z}_A^\dagger]\, | \psi\rangle  = -\epsilon\langle\phi|\psi\rangle,
\ee
where we recall that ${\cal Z}_A = {\cal Q}(Z_A)$.  The right hand side of (\ref{f1.3}) is simply
\be\label{f1.1}
-\epsilon\langle\phi|\psi\rangle = -\epsilon\int_\Sigma \bar\phi\psi\,\Omega.
\ee
The left hand side of (\ref{f1.3}), on the other hand, requires some
manipulations and is given by
\bea
\langle \phi |\, [{\cal Z}_A,{\cal Z}_A^\dagger]\, | \psi\rangle &=&
\langle \phi |\, {\cal Q}(Z_A) {\cal Q}(\bar Z_A) - {\cal Q}(\bar Z_A){\cal Q}(Z_A)    \, | \psi\rangle\nonumber\\
 &=& \langle \phi |\, \pi Z_A \pi \bar Z_A   \, | \psi\rangle-
\langle \phi |\,\pi \bar Z_A \pi Z_A  \, | \psi\rangle\nonumber\\
&=& \langle \phi |\, Z_A \pi \bar Z_A   \, | \psi\rangle-
\langle \phi |\,\bar Z_A Z_A  \, | \psi\rangle,\nonumber
\eea
where we have used the properties of $\cal Q$ described at the beginning of this section.
We may now use the expansion (\ref{bk3.1}) for the action of the Bergman projection and write
\bea
&&\langle \phi |\, [{\cal Z}_A,{\cal Z}_A^\dagger]\, | \psi\rangle =\nonumber\\
&&= \sum_{n=0}^\infty (-1)^n 
\int_\Sigma  \bar\phi Z_A \left( \nabla {1\over P_1}\dots\nabla {1\over P_n} P_1\dots P_n
{1\over P_n}\bar\nabla\dots{1\over P_1}\bar\nabla \bar Z_A\psi\right)\,\Omega - 
\int_\Sigma  \bar \phi \bar Z_A Z_A\psi\,\Omega \nonumber\\
&&= \sum_{n=1}^\infty (-1)^n 
\int_\Sigma  \bar\phi Z_A \left(\nabla {1\over P_1}\dots\nabla {1\over P_n} P_1\dots P_n
{1\over P_n}\bar\nabla\dots{1\over P_1}\bar\nabla \bar Z_A\psi\right)\,\Omega.  \nonumber
\eea
We can then integrate by parts to conclude that
\bea\label{f1.2}
%&&\langle \phi |\, [{\cal Z}_A,{\cal Z}_A^\dagger]\, | \psi\rangle
%=\\
\langle \phi |\, [{\cal Z}_A,{\cal Z}_A^\dagger]\, | \psi\rangle
&= &\sum_{n=1}^\infty  
\int_\Sigma  P_1\dots P_n \left({1\over P_n}\nabla \dots {1\over P_1}\nabla \bar\phi Z_A\right)  
\left( {1\over P_n}\bar\nabla\dots{1\over P_1}\bar\nabla \bar Z_A\psi\right)\,\Omega \nonumber\\
&= &\sum_{n=1}^\infty
\int_\Sigma  \bar\phi \psi\, (P_1\dots P_n) \left({1\over P_n}\nabla\dots {1\over P_1}\nabla  Z_A\right)  
\left( {1\over P_n}\bar\nabla\dots{1\over P_1}\bar\nabla \bar Z_A\right)\,\Omega, \nonumber
\eea
where, in the last line, we have used the fact that $\bar\phi$ and
$\psi$ are respectively antiholomorphic and holomorphic and therefore
commute with the covariant derivatives $\nabla$ and $\bar\nabla$.
Note that in (\ref{f1.2}) the covariant derivatives $\nabla$ act only
on the terms within the first parentheses; for the remainder of the
paper we drop our convention that covariant derivatives act on all
terms to their right, and instead take covariant derivatives to act on
all terms to their right within a given parenthetical term.
Comparing (\ref{f1.1}) and (\ref{f1.2}) we conclude that equation
(\ref{f1.3}) is satisfied if
\be\label{f1.4}
-\epsilon= \sum_{n=1}^\infty P_1\dots P_n \left({1\over P_n}\nabla\dots {1\over P_1}\nabla  Z_A\right)  
\left( {1\over P_n}\bar\nabla\dots{1\over P_1}\bar\nabla \bar Z_A\right).
\ee
Note that, in the above equation, all the covariant
derivatives can be replaced by partial derivatives ($\nabla$ acts on
tensors of holomorphic dimension $0$, $\bar\nabla$ on tensors of
antiholomorphic dimension $0$).  Therefore equation (\ref{f1.4}) does
not contain the connection explicitly and can be considered as an
equation for the curvature tensor $R$.  

We have now found a set of equations which in principle could be used to
construct a matrix representation of an arbitrary holomorphic curve.
Given a set of embedding functions $Z_A$, we need to solve equations
(\ref{f1.4}) and (\ref{bk3.2}) for the tensors $P_n$.  In principle
we can then
use $P_1 = R$ to determine $\Omega$, which fixes the inner product on
${\cal H}$ and gives a matrix representation of the holomorphic curve
defined by the $Z_A$'s.  Unfortunately, however, we do not have any
way of exactly solving these equations in general.  Thus, we now
describe a perturbative solution of the problem.

We wish to solve (\ref{f1.4})
perturbatively in $\epsilon$ and, to this end, we recall the planar
membrane example of section \ref{m3}.  In that case we noted that the
correct integration measure $\Omega$ was given by $\Omega\propto
e^{-z\bar z/\epsilon} dz\wedge d\bar z$, which is the case analyzed in
section \ref{sec:bk-example}.  In fact, if one takes $Z_1=z$,
$Z_A=0\,(A=2,3,4)$ and $P_n=-n/\epsilon$, one can quickly verify that
equation (\ref{f1.4}) is satisfied (only the $n=1$ term is non-zero).
The important fact to be learned from this example is that $P_n
\sim 1/\epsilon$.  In the general case we then define
$$P_n= {1\over\epsilon} Q_n,$$ where, as we will see shortly, the
$Q_n$'s are analytic in $\epsilon$ and can therefore be computed
perturbatively.  Equation (\ref{f1.4}) can now be written as
\bea\label{f1.5}
-1 &=& \sum_{n=1}^\infty \epsilon^{n-1} \pi_n\\
\pi_n &=& Q_1\dots Q_n \left({1\over Q_n}\partial\dots {1\over
Q_1}\partial  Z_A\right)   
\left( {1\over Q_n}\bar\partial\dots{1\over Q_1}\bar\partial \bar
Z_A\right).\nonumber 
\eea
Let us show that the above equation can be solved for the $Q_n$'s
order by order in powers of $\epsilon$.  First we note that the
tensors $Q_n$ satisfy the recursion relation
\bea\label{f1.8}
Q_1 &=& \epsilon R\nonumber\\
Q_n &=& Q_{n-1} + Q_1 + \epsilon\sum_{j=1}^{n-1} \partial\bar\partial\ \ln\ Q_j,
\eea
which is an immediate consequence of equation (\ref{bk3.2}).
Therefore one needs only to solve for $Q_1$ at any given order and the
rest of the $Q_n$'s are automatically determined by (\ref{f1.8}).  To
solve for $Q_1$ we assume that all the $Q_n$'s are known to order
$\epsilon^{N-1}$.  From equation (\ref{f1.5}) we see that also the
$\pi_n$'s are known to the same order.  We now wish to compute $Q_1$
to order $\epsilon^N$.  To this end we rewrite equation (\ref{f1.5})
as
\be\label{f1.6}
-Q_1 = \partial Z_A \bar\partial \bar Z_A + \epsilon\left( \sum_{n=2}^{N+1} \epsilon^{n-2}\pi_n Q_1\right) + 
{\cal O} (\epsilon^{N+1})
\ee
and we note that the term in parentheses multiplies $\epsilon$ and
therefore needs only to be computed to order $\epsilon^{N-1}$.  But
both $\pi_n$ and $Q_1$ are known to order $\epsilon^{N-1}$ by
assumption and therefore equation (\ref{f1.6}) determines $Q_1$ to
order $\epsilon^N$.  Let us note that this procedure is completely
algebraic and that no partial differential equation needs to be
solved.

We have solved, at least perturbatively, for $Q_1$.  Our real goal, on the other hand, is to determine
the integration $2$-form $\Omega$.  To this end we fix on the surface $\Sigma$ a  holomorphic
$(1,0)$-form $$H=h\,dz$$
We then rewrite
\bea
\Omega &=& iH\wedge \bar H\,e^{-{\cal K}/\epsilon}\nonumber\\
&=& ih\bar h \,e^{-{\cal K}/\epsilon}\,dz\wedge d\bar z,\nonumber
\eea
where $\cal K$ is a real scalar function on the surface (we have used
the fact that $\Omega$ is real and positive).  Recalling that
$\Omega=iC\,dz\wedge d\bar z$ and that
$R=\partial\bar\partial\,\ln\,C=(1/
\epsilon) Q_1$, we conclude that
\be\label{f1.7}
-Q_1 = \partial\bar\partial {\cal K}
\ee
(we have used the fact that $\partial\bar\partial\,\ln(h\bar h)=0$).  This shows that, as long as we 
succeed in writing $Q_1$ as the laplacian of a scalar function $\cal K$, we have solved the problem 
completely.

Let us first analyze the solution of equations (\ref{f1.5}) and
(\ref{f1.7}) in the $\epsilon\rightarrow 0$ classical limit.  In this
case equation (\ref{f1.6}) reduces to $$-Q_1 = \partial Z_A
\bar\partial \bar Z_A +{\cal O} (\epsilon)$$ and therefore $${\cal
K}=Z_A\bar Z_A + {\cal O} (\epsilon).$$ In this limit we then have
\be\label{f1.21}
\Omega \simeq H\wedge\bar H\,e^{-Z_A\bar
Z_A/\epsilon}.\hskip1cm(\epsilon\rightarrow 0\  {\rm classical\ limit}) 
\ee 
We will see below that the above result is what is expected if one
uses geometric quantization techniques. We note that, for $\epsilon\rightarrow 0$, $R=-(1/\pi)\mu$. The 
curvature of the measure $\Omega$ gives, in this limit, the density of $0$-branes.

One can go on and compute higher order corrections to $Q_1$ and therefore to $\cal K$.  A long but mechanical computation,
following the procedure outlined above, shows that, to order $\epsilon^2$
\be\label{f1.20}
{\cal K} = Z_A\bar Z_A - {\epsilon\over 2} \ln\left({\alpha\over h\bar h}\right)+{\epsilon^2\over 6}{1\over\alpha}\partial
\bar\partial\,\ln\,\alpha+ \dots,
\ee
where $$\alpha=\partial Z_A \bar\partial Z_A.$$ Only one point about the derivation of equation (\ref{f1.20}) is
worth mentioning.  Using the procedure described previously, one computes $-Q_1 = \dots -(\epsilon/2)\partial\bar
\partial\,\ln\,\alpha+\dots$, so that one is tempted to set ${\cal K} = \dots - (\epsilon/2)\,\ln\,\alpha$.  This is wrong,
since $\alpha$ is a $(1,1)$ tensor, and therefore $\ln\,\alpha$ is not
a scalar.  On the other hand $\ln(\alpha/h\bar h)$ is a scalar, and
moreover $\partial\bar\partial\,\ln(\alpha/h\bar h) =
\partial\bar\partial\,\ln\,\alpha$.  Equation (\ref{f1.20}) implies
that the first quantum correction of equation (\ref{f1.21}) is given
by
\be\label{f1.22}
\Omega\simeq dz\wedge d\bar z\,\sqrt{h\bar h} \sqrt{\partial Z_A \bar\partial Z_A}\, e^{-Z_A\bar Z_A/\epsilon}.
\ee
We will see in the next section that this corresponds to the metaplectic correction to geometric quantization.

Finally we note that $\cal K$ is not uniquely defined by equation
(\ref{f1.7}).  In particular, if $\chi$ is a holomorphic function, the
transformation $${\cal K}\rightarrow {\cal
K}-\epsilon\chi-\epsilon\bar\chi$$ leaves $Q_1$ invariant.  This
transformation corresponds to the one analyzed in section
\ref{sec:bk-properties} since $\Omega\rightarrow\Omega\,
e^{\chi+\bar\chi}$.

We conclude this section with a brief synopsis of our solution to the
problem posed in section 2.3.  We have found a system of equations
(\ref{f1.4}, \ref{bk3.2}) whose solution in principle determines a
matrix representation of any holomorphic curve.  We could not solve
this system of equations in full generality, but found a perturbative
solution by expanding in the parameter $\epsilon$.  To find the
perturbative solution to order $\epsilon^n$ it is necessary to
solve the system of equations (\ref{f1.5}, \ref{f1.8}) to order
$\epsilon^n$.  The relation (\ref{f1.7}) can then be used to construct
${\cal K}$; the general form of ${\cal K}$ is given to order
$\epsilon^2$ in (\ref{f1.20}).  Once ${\cal K}$ is known, the
corresponding $\Omega$ can be used to explicitly construct the
matrices $X_i$ for the holomorphic curve in question.

%%%%%%%%%%%%%%%%%%%%%%%%%%%%%%%%%%%%%%%%%%%%%%%%%%%%%%%%%%%%%%%%%%%%%%%%%
\section{Numerical Interlude}
\label{n}
%%%%%%%%%%%%%%%%%%%%%%%%%%%%%%%%%%%%%%%%%%%%%%%%%%%%%%%%%%%%%%%%%%%%%%%%%

In this section we use the results of the last section 
and apply them to two numerical examples.

First we analyze once more the parabolic membrane example of section \ref{m3}.
We are going to use the notation of that section.  We recall that $\cal
H$ is spanned by the orthogonal states $|v_n\rangle=|z^n\rangle$, with inner product
$\langle v_n|v_n\rangle=k_n$.  In particular (choosing $H=dz$)
\be\label{n.1}
k_n= \int dz\wedge d\bar z\,z^n\bar z^n e^{-{\cal K}/\epsilon}.
\ee
We also recall that $\alpha_n=k_{n+1}/k_n$ and that $\alpha_0$ could
be computed to arbitrary precision.  We can now use the successive
approximations (\ref{f1.20}) of $\cal K$ and compute $\alpha_0$
numerically using the integral (\ref{n.1}).  The results are shown in
figure 1, together with the exact numerical result computed following
the prescription given in section \ref{m3}.

\begin{center}
\leavevmode
\epsffile{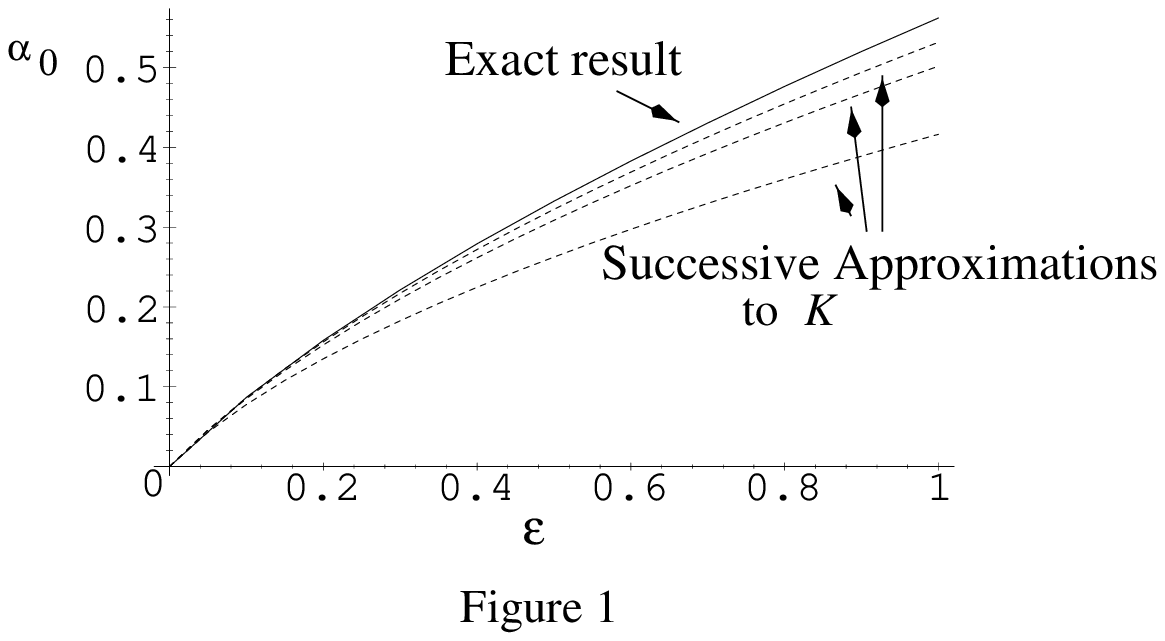}
\end{center}

As a second example we consider a curve that cannot be analyzed with the techniques
of section \ref{m3}. To be specific we look at the curve defined by
$$
Z_1^2 = Z_2+ 2 Z_2^2 + Z_2^3
$$
and parametrized by
\bea
Z_1 &=& z+z^3\nonumber\\
Z_2 &=& z^2. \nonumber
\eea
The space $\cal H$ is still spanned by the vectors $|v_n\rangle = |z^n\rangle$, but the basis 
$|v_n\rangle$ is not orthonormal (it is not even orthogonal). If we let
$$I_{nm} = \langle v_n | v_m \rangle = \int dz\wedge d\bar z\  \bar z^n z^m\ e^{-{\cal K}/\epsilon}$$
be the inner-product matrix, we can move to an orthonormal basis by diagonalizing $I$ with a
matrix $D$ such that $D^\dagger I D = {\rm Id}$. Any operator on $\cal H$ which is represented
by a matrix $A$ in the basis $|v_n\rangle$ is represented by $D^{-1} A D$ with respect to the orthonormal
basis. In particular the matrices ${\cal Z}_1$ and ${\cal Z}_2$ are given by

\[
{\cal Z}_1 =D^{-1}\left({\small \begin{array}{cccccc}
0 & 0 & 0 & 0 & \ddots & \\
1 & 0 & 0 & 0 & 0      & \ddots\\
0 & 1 & 0 & 0 & 0      & \ddots \\
1 & 0 & 1 & 0 & 0      & \ddots\\
\ddots & 1 & 0 & 1 & 0 & \ddots\\
& \ddots & \ddots & \ddots &\ddots & \ddots\\
\end{array}}\right) D,
\;\;\;\;\;
{\cal Z}_2 =D^{-1}\left({\small\begin{array}{cccccc}
0 & 0 & 0 & 0 & \ddots & \\
0 & 0 & 0 & 0 & 0      & \ddots\\
1 & 0 & 0 & 0 & 0      & \ddots \\
0 & 1 & 0 & 0 & 0      & \ddots\\
\ddots & 0 & 1 & 0 & 0 & \ddots\\
& \ddots & \ddots & \ddots &\ddots & \ddots\\
\end{array}}\right) D
\]
and satisfy (by construction) 
$$
{\cal Z}_1^2 = {\cal Z}_2+ 2 {\cal Z}_2^2 + {\cal Z}_2^3.
$$
In order to find a numerical approximation to the matrices ${\cal Z}_1$ and ${\cal Z}_2$ we first
of all take $\epsilon$ small ($0.001$), so that we can approximate ${\cal K} \simeq Z_A \bar Z_A$.
Moreover we restrict ourselves to finite $N\times N$ matrices (more precisely, we restrict to the
subspace of $\cal H$ spanned by the vectors $|v_n\rangle$, $n<N$). If we take $N$ large enough ($N=10$)
we expect that the corresponding matrices ${\cal Z}_A$ will be accurate, at least in the 
upper-left corner (physically this corresponds to the region of the brane with $Z_1$, $Z_2$ small). 
The matrices $I$ and $D$ can then be computed numerically and can be used to evaluate the coordinate
matrices. The result is

\[
{\cal Z}_1 \approx {1\over 1000} \left({\small \begin{array}{cccccc}
0 & -0.1 & 0 & 0 &  0      & \\
31+1.9i & 0 & -0.3 & 0 & 0      & \ddots\\
0 & 45-2.6i & 0 & -0.5 & 0      & \\
-0.1 & 0 & 54-6.7i & 0 & -0.7      & \ddots\\
0  & 0.15  & 0 & 63+7.7i & 0 & \\
& \ddots & & \ddots & & \ddots\\
\end{array}}\right) + {\cal O}(10^{-5})
\]
and

\[
{\cal Z}_2 \approx {1\over 1000} \left({\small \begin{array}{cccccc}
0 & 0 & 0 & 0 &  0      & \\
0 & 0 & 0 & 0 & 0      & \ddots\\
1.4 & 0 & 0 & 0 & 0      & \\
0 & 2.4-0.4i & 0  & 0 & 0      & \ddots\\
0  & 0  & 3.4 & 0 & -0.1 & \\
& \ddots & & \ddots & & \ddots\\
\end{array}}\right) + {\cal O}(10^{-5})
\]
The above matrices satisfy, as they should,
$$[{\cal Z}_A, {\cal Z}_A^\dagger]=-\epsilon \;{\rm Id}+{\cal O}(10^{-5}).$$

%%%%%%%%%%%%%%%%%%%%%%%%%%%%%%%%%%%%%%%%%%%%%%%%%%%%%%%%%%%%%%%%%%%%%%%%%
\section{Comparison with Geometric Quantization}
\label{f2}
%%%%%%%%%%%%%%%%%%%%%%%%%%%%%%%%%%%%%%%%%%%%%%%%%%%%%%%%%%%%%%%%%%%%%%%%%

In this final section we analyze the connection between our results and the theory of geometric quantization \cite{Woodhouse,Tuynman,Tuynman2}.

We recall briefly a few key concepts of prequantization.  We let
$(\Sigma,\omega)$ be a symplectic manifold of dimension $2n$
\footnote{$\omega$ is a closed non-degenerate $2$-form.  Given a
function $A$, the corresponding hamiltonian flow vector field $X_A$ is
defined by $dA(\cdot)+\omega(X_A,\cdot)=0.$ The Poisson bracket of two
functions is given by $\{ A,B\} =X_A(B) = \omega(X_A,X_B)$ and it
satisfies $[X_A,X_B]=X_{\{ A,B\} }$}.  If $\omega$ is integral on
closed $2$-cycles (quantization condition), one may consider
$2\pi\omega$ as the field strength of a $U(1)$ gauge potential.
Locally $\omega=d\theta$, where $2\pi\theta$ is the gauge potential.
The covariant derivative is then $$\nabla=d-2\pi i\theta.$$ In
prequantization one considers the Hilbert space $\cal V$ of sections
of the above $U(1)$ bundle with inner product $$\langle \eta
|\xi\rangle = \int_\Sigma\omega^n\,\bar\eta\xi,$$ and assigns to
functions $A$ on $\Sigma$ the first order differential operator
$${\cal Q}(A) = -{i\over 2\pi}\nabla_{X_A} + A,$$ where $\nabla_{X_A}$
denotes the covariant derivative in the direction of the hamiltonian
flow $X_A$ of $A$.  Using the fact that $[\nabla_X,
\nabla_Y]=\nabla_{[X,Y]} -2\pi i \omega(X,Y)$, one can show that
$${\cal Q}(\{ A,B \}) = 2\pi i[{\cal Q}(A), {\cal Q}(B)].$$

This completes the description of the prequantization.  To complete
the geometrical quantization process it is necessary to choose a
polarization.  There is a particularly simple  choice of
polarization when our symplectic manifold is a Riemann surface.  
In this case it is natural to choose a polarization given by the condition
that the sections $\eta$ be holomorphic.  
We now assume that $\Sigma$ is a Riemann surface.  Using the Dolbeault
lemma (or considering $\omega$ as a K\"{a}hler form), we can find a
cover ${\cal U}_i$ of $\Sigma$ and real functions ${\cal K}_i$ on
${\cal U}_i$ such that $$\omega = -{1\over 2\pi i
\epsilon}\partial\bar\partial\,{\cal K}_i.$$ We also have that ${\cal
K}_j = {\cal K}_i + \chi_{ij}+\bar\chi_{ij}$, for some $\chi_{ij}$
holomorphic on ${\cal U}_i\cap{\cal U}_j$.  The $U(1)$ potential on
${\cal U}_i$ is given by $$\theta_i = -{1\over 4\pi
i\epsilon}(\bar\partial {\cal K}_i-\partial {\cal K}_i),$$ and one can
check that $\theta_j=\theta_i+ d\alpha_{ij}$, where
$\alpha_{ij}={1\over 4\pi i\epsilon}(\chi_{ij}-\bar\chi_{ij})$.
Sections of the $U(1)$ bundle are given by functions $\eta_i$ on
${\cal U}_i$ related by $\eta_j=\eta_i e^{2\pi i\alpha_{ij}}$.  
A holomorphic section $\eta$ satisfies the equation
$$\bar\nabla
\eta_i = \bar\partial\eta_i + {1\over
2\epsilon}\eta_i\bar\partial{\cal K}_i = 0$$ which is solved by $$\eta_i =
\phi_i e^{-{\cal K}_i/2\epsilon},$$ with $\phi_i$ holomorphic.  One
can check that $\phi_j=\phi_i e^{\chi_{ij}/\epsilon}$ and therefore
the definition is consistent.  One may then restrict the attention
from $\cal V$ to the space $\cal H$ of holomorphic sections of the
holomorphic line bundle $L$ with transition functions $\lambda_{ij}=
e^{\chi_{ij}/\epsilon}$.  The inner product on $\cal H$ is given by
$$\langle\phi|\psi\rangle=\int_\Sigma\bar\phi\psi\,e^{-{\cal
K}/\epsilon}\,\omega.$$ First we note that, if $A$ is a holomorphic
function, then $X_A$ is a $(0,-1)$ vector field and therefore
$\nabla_{X_A} \propto \bar\nabla$.  This means that, on holomorphic
sections, ${\cal Q}(A) = A$ as we assumed in the paper.  Moreover we
note that, in our specific case, $\omega = \mu\,d^2\sigma = -{1\over
2\pi i\epsilon}
\partial\bar\partial\, Z_A\bar Z_A$, so that the line bundle $L$ is trivial with globally defined K\"{a}hler
potential $${\cal K}= Z_A\bar Z_A.$$ The space $\cal H$ can then be identified with the space of holomorphic functions
 with inner product given by
 \bea
 \langle\phi|\psi\rangle &=& \int_\Sigma\bar\phi\psi\,\Omega\nonumber\\
  \Omega &=& e^{-{Z_A\bar Z_A}/\epsilon}\,\omega.\nonumber
  \eea 
  We see that
 we recover equation (\ref{f1.21}), with $\omega$ replacing $H\wedge\bar H$.  On the other hand this difference is irrelevant
 in the $\epsilon\rightarrow 0$ limit, as can be checked by computing the leading $1/\epsilon$ behavior of $R$ in both
 cases.
 
 Standard geometric quantization is improved by the metaplectic correction \cite{Tuynman2}, which we now describe (as it reads
 in our present setting).  One assumes that the ${\cal U}_i$ are coordinate patches with local coordinate $z_i$ and
 considers the holomorphic line bundle $K$ with transition functions given by $\kappa_{ij} = \sqrt{{\partial z_i\over
 \partial z_j}}$ \footnote{This is the square root of the canonical bundle.  
 Its existence poses problems similar to the ones encountered in choosing a spin structure,
 since $\sqrt{\,}$ is defined only up to a sign}.  One then considers, as Hilbert space $\cal H$, the holomorphic sections
 of the tensor bundle $K\otimes L$, given by analytic functions $\eta_i$ related by $\eta_j= \lambda_{ij} \kappa_{ij}
 \eta_i$.  To construct an inner product between two sections $\eta$ and $\xi$
 one has to build a scalar from $\bar\eta_i$ and $\xi_i$.  We recall that
 $\omega={i\over 2} \mu_i dz_i\wedge d\bar z_i$, where 
 $\mu_j= \kappa^2_{ij}\bar\kappa^2_{ij}\mu_i$.  This allows us to construct a scalar, since $\mu_i^{-1/2}\bar\eta_i\xi_i
 e^{-{\cal K}_i/\epsilon}$ is invariant if we change $i\rightarrow j$.  If one considers the vector field $X_i=X_{z_i} = {-2i
 \over\mu_i}\bar\partial_i$, we see that ${1\over 2\pi i\epsilon} \omega(\bar X_i,X_i) 
  = (\pi\epsilon\mu_i)^{-1}$.  Therefore the scalar we where looking for is $\sqrt{(2\pi i\epsilon)^{-1}
 \omega(\bar X,X)}\,\bar\eta\xi\, e^{-{\cal K}/\epsilon}$.  The inner product is then given by 
 $$\langle\eta|\xi\rangle=\int_\Sigma \omega\,\sqrt{{1\over 2\pi
 i\epsilon} \omega(\bar X,X)}\,\bar\eta\xi\, e^{-{\cal K}/\epsilon}.$$
 We now specialize to our specific case.  First of all we recall that
 $ \pi\epsilon\mu_i=\partial_i Z_A \bar\partial_i \bar Z_A$.  We have
 seen that the bundle $L$ is trivial.  We assume that we can also
 trivialize the bundle $K$, and we consider a nowhere zero section
 $f_i$ of $K$.  First we note that $h_i=f_i^2$ is a section of the
 canonical line bundle, and therefore $H=h_i\,dz_i$ is well defined
 holomorphic one-form.  Moreover any section $\eta$ of $K\otimes L$
 can be written as $\eta_i = f_i \phi=\sqrt{h_i} \phi$, where $\phi$
 is a globally defined analytic function.  This means that $\cal H$
 can then be identified with the space of holomorphic functions with
 inner product given by \bea \langle\phi|\psi\rangle &=&
 \int_\Sigma\bar\phi\psi\,\Omega\nonumber\\ \Omega &=& -{1\over 2\pi i
 \epsilon} dz\wedge d\bar z \,\sqrt{h\bar h} \sqrt{\partial Z_A
 \bar\partial \bar Z_A}\, e^{-{Z_A\bar Z_A}/\epsilon}.\nonumber \eea
 We then recover equation (\ref{f1.22}) of section \ref{f1} (up to an
 irrelevant scale factor).
 
 We conclude this section by analyzing once more the $\epsilon \rightarrow 0$ limit of the quantization
 prescription analyzed in section \ref{f1}.  First of all we notice that, given functions $A$ and $B$ on 
 $\Sigma$ and elements $\phi,\psi\in {\cal H}$, one has that 
 $\langle \phi |\, {\cal Q}(A) {\cal Q}(B) | \psi\rangle=
 \langle \phi |\, \pi A\pi B   \, | \psi\rangle = 
\langle \phi |\,A\pi B\, | \psi\rangle$.  Following the same steps as in section \ref{f1}, one can then show that
$$\langle \phi |\, {\cal Q}(A) {\cal Q}(B) | \psi\rangle =
\langle \phi |\, {\cal Q}(A \star B) \, | \psi\rangle,$$
where the $\star$ product is defined by
$$
A\star B = \sum_{n=0}^\infty \epsilon^n Q_1\dots Q_n \left({1\over Q_n}\partial\dots {1\over Q_1}\partial  A\right)  
\left( {1\over Q_n}\bar\partial\dots{1\over Q_1}\bar\partial B\right).
$$
The product $\star$ is not commutative.  Moreover one has that $A\star B = AB + (\epsilon/Q_1) \cdot \partial A\bar\partial B+\dots$,
so that one can derive the important equations valid in the $\epsilon\rightarrow 0$ limit
\bea
A\star B &=& AB + \dots\nonumber\\
A\star B- B\star A &=& {1\over 2\pi i} \{ A,B \} +\dots,\nonumber
\eea
where we have used that $Q_1=-\partial Z_A\bar\partial \bar Z_A +\dots$.

%%%%%%%%%%%%%%%%%%%%%%%%%%%%%%%%%%%%%%%%%%%%%%%%%%%%%%%%%%%%%%%%%%%%%%%%%
\section{Conclusion}
\label{c}
%%%%%%%%%%%%%%%%%%%%%%%%%%%%%%%%%%%%%%%%%%%%%%%%%%%%%%%%%%%%%%%%%%%%%%%%%

We have analyzed the problem of constructing matrix representations of
holomorphic curves describing static membranes in M-theory.  We
discussed some simple examples with rotational symmetry for which the
corresponding matrices were easy to construct explicitly.  We also
developed a more general approach which gives the matrix
representation of an arbitrary holomorphic curve.  We were unable to
exactly solve the equations in the general case; we showed, however,
that these equations can be solved perturbatively in the parameter
$\epsilon$.  This parameter is related to the inverse density of
0-branes on the membrane and goes to 0 in the limit of a smooth
membrane.  The $\epsilon$ expansion is expected to be an asymptotic
series for the exact expressions, and should give a good approximation
to a matrix representation of any holomorphic curve.  The first two
terms in the $\epsilon$ expansion correspond to the standard result
from geometric quantization and the metaplectic correction to this
result.

We conclude this paper by suggesting some future directions for
research.

1) First of all, from a practical point of view, it would be very
useful to obtain a recursive formula for the higher corrections to the
function $\cal K$.  We have not been able to derive a closed form
expression for $\cal K$, but we believe that, with some work, it
should be possible to extract it from equation (\ref{f1.4}).

%2) The matrices $\cal Z_A$ represent classical static solutions to the
%matrix theory equations of motion.  It would be desirable to
%understand to what extent their stability is effected by quantum
%effects, and to this end an initial step would be to compute the first
%quantum correction, coming from the one-graviton exchange diagrams.

2) The holomorphic membranes we have considered here are stable BPS
configurations which preserve some of the supersymmetries of the
theory.  It would be interesting to understand the physical properties
of these solutions better, either in supergravity or in matrix theory.
In particular, to the best of our knowledge the supergravity solutions
corresponding to these holomorphic membrane configurations are not
known.  It would be nice to have an explicit construction of these
solutions.  From the matrix theory point of view, it would be nice to
explicitly demonstrate the supersymmetry of these configurations.

3) Finally, one should analyze the problem of holomorphic curves
embedded in compactified space.  If the formalism can be generalized,
one can apply it, in particular, to the case of compactification on
$T^4$.  In this case the defining equations (\ref{m2.6}) and
(\ref{m2.8}) are, after $T$-duality, the equations of self-duality of
the dual Yang-Mills field.  In particular one can try to understand
the relation between holomorphic curves embedded in $T^4$ and
Yang-Mills instantons on the dual $T^4$.

\section*{Acknowledgements}

We would like to thank
Larus Thorlacius and Dan Waldram for helpful discussions.
The work of WT is supported in part by the National Science Foundation
(NSF) under contract PHY96-00258.

\newpage

%\bibliography{branes}

\begin{thebibliography}{10}

\bibitem{BFSS}
T.\ Banks, W.\ Fischler, S.\ Shenker, and L.\ Susskind, ``M Theory as a Matrix
  Model: A Conjecture,'' \PR {\bf D55} (1997) 5112, {\tt hep-th/9610043}.

\bibitem{Bilal-review}
A.\ Bilal, ``M(atrix) theory: a pedagogical introduction,'' {\tt
  hep-th/9710136}.

\bibitem{Banks-review}
T.\ Banks, ``Matrix Theory,'' \NP {\it Proc. Suppl.} {\bf 67} (1998) 180, {\tt
  hep-th/9710231}.

\bibitem{Susskind-review}
D.\ Bigatti and L.\ Susskind, ``Review of matrix theory,'' {\tt
  hep-th/9712072}.

\bibitem{WT-Trieste}
W.\ Taylor, ``Lectures on D-branes, gauge theory and M(atrices),'' Proceedings
  of Trieste summer school 1997, to appear; {\tt hep-th/9801182}.

\bibitem{Goldstone-Hoppe}
J.\ Goldstone, unpublished; J.\ Hoppe, MIT Ph.D.\ thesis (1982); J.\ Hoppe, in
  proc.\ Int.\ Workshop on Constraint's Theory and Relativistic Dynamics; eds.
  G.\ Longhi and L.\ Lusanna (World Scientific, 1987).

\bibitem{dhn}
B.\ de Wit, J.\ Hoppe and H.\ Nicolai, \NP {\bf B305} [FS 23] (1988) 545.

\bibitem{Dan-Wati}
D.\ Kabat and W.\ Taylor, ``Spherical membranes in Matrix theory,'' {\em Adv.
  Theor. Math. Phys.} {\bf 2}, 181-206, {\tt hep-th/9711078}.

\bibitem{bss}
T.\ Banks, N.\ Seiberg, and S.\ Shenker, ``Branes from Matrices,'' {\tt
  hep-th/9612157}.

\bibitem{Woodhouse}
N.\ M.\ J.\ Woodhouse, {\em Geometric Quantization} (2nd Ed.), (Oxford\ Univ.\
  Press).

\bibitem{bergman}
S.\ Bergman, {\em Kernel function and conformal mapping}, (2nd rev.\ ed.) (American
  Mathematical Society, 1970).

\bibitem{Tuynman}
G.\ M.\ Tuynman, ``Generalized Bergman kernels and geometric quantization,''
  {{\em J.\ Math.\ Phys.\ }} {\bf 28} (1987) 3 .

\bibitem{Tuynman2}
G.\ M.\ Tuynman, ``Quantization: Towards a comparison between methods,'' {{\em
  J.\ Math.\ Phys.\ }} {\bf 28} (1987) 12 .

\bibitem{Susskind-DLCQ}
L.\ Susskind, ``Another Conjecture about M(atrix) Theory,'' {\tt
  hep-th/9704080}.

\bibitem{Sen-DLCQ}
A.\ Sen, ``D0 Branes on $T^n$ and Matrix Theory,'' {\tt hep-th/9709220}.

\bibitem{Seiberg-DLCQ}
N.\ Seiberg, ``Why is the Matrix Model Correct?,'' \PRL {\bf 79} (1997) 3577,
  {\tt hep-th/9710009}.

\end{thebibliography}
\bibliographystyle{plain}

\end{document}